%% file: main.tex
\documentclass[conference]{commons/IEEEtran}
\input{commons/pkgs.tex}
\input{commons/commands.tex}

\begin{document}
\input{0_title.tex}
\input{authors.tex}

\maketitle
\thispagestyle{plain}
\pagestyle{plain}
\begin{abstract}
    \input{1_abstract}
\end{abstract}
\input{0_main.tex}
\bibliographystyle{IEEEtran}
\bibliography{0_ref}
\appendices
\input{9_appendix}

\end{document}

%% file: commons/pkgs.tex
\usepackage{CJKutf8}
\usepackage{amsmath}
\usepackage{graphicx}
\usepackage[tight,footnotesize]{subfigure}
\usepackage{multirow}
\usepackage{fancyvrb}
\usepackage[flushleft, online]{threeparttable}
\usepackage{xurl}
\usepackage[
    colorlinks,
    linkcolor=blue,
    anchorcolor=blue,
    citecolor=blue,
    urlcolor=blue
]{hyperref}
\usepackage{array}
\usepackage{algorithmic}
\usepackage{booktabs}
\usepackage{xcolor}
\usepackage{ulem}
\usepackage{authblk}

%% file: commons/commands.tex
\newcommand{\subject}[1]{\vspace{3pt}\noindent\textbf{#1}}
\newcommand{\subsubject}[1]{\vspace{3pt}\noindent\textit{#1}}

\newcommand\malurl[1]{\href{notalink}{{\nolinkurl{#1}}}}

\newcounter{finding}
\newcommand{\finding}[1]{
\vspace{3pt}
\noindent
\framebox{
\begin{minipage}[b]{.95\columnwidth}
\noindent \textbf{Finding \Roman{finding}}: \textit{#1}
\stepcounter{finding}
\end{minipage}
}
\vspace{-8pt}
}

\newcommand{\ignore}[1]{}

%% file: 0_title.tex
\title{Enabling Privacy-Preserving Cyber Threat Detection with Federated Learning}

%% file: authors.tex
\author[1]{Yu Bi}
\author[1]{Yekai Li}
\author[2]{Xuan Feng}
\author[*,1]{Xianghang Mi\thanks{*Corresponding author.}}

\affil[1]{University of Science and Technology of China}
\affil[2]{Microsoft Research Asia}

\renewcommand*{\Affilfont}{\small\it}
\IEEEoverridecommandlockouts

%% file: 1_abstract.tex
Despite achieving good performance and wide adoption, machine learning based security detection models (e.g., malware classifiers) are subject to concept drift and evasive evolution of attackers, which renders up-to-date threat data as a necessity.  However, due to enforcement of various privacy protection regulations (e.g., GDPR),   it is becoming increasingly challenging or even prohibitive for security vendors to collect individual-relevant and privacy-sensitive threat datasets, e.g., SMS spam/non-spam messages from mobile devices. To address such obstacles, this study systematically profiles the (in)feasibility of federated learning for privacy-preserving cyber threat detection in terms of effectiveness, byzantine resilience, and efficiency. This is made possible by the build-up of multiple threat datasets and threat detection models, and more importantly, the design of realistic and security-specific experiments. 

We evaluate FL on two representative threat detection tasks, namely SMS spam detection and Android malware detection. It shows that FL-trained detection models can achieve a performance that is comparable to centrally trained counterparts. Also, most non-IID data distributions  have either minor or negligible impact on the model performance, while a label-based non-IID distribution of a high extent can incur non-negligible fluctuation and delay in FL training. Then, under a realistic threat model, FL turns out to be adversary-resistant to attacks of both data poisoning and model poisoning. Particularly, the attacking impact of a practical data poisoning attack is no more than 0.14\% loss in model accuracy. Regarding FL efficiency,  a bootstrapping strategy turns out to be  effective to mitigate the training delay as observed in label-based non-IID scenarios.

%% file: 0_main.tex
\section{Introduction}
\label{sec:intro}
\input{2_intro.tex}

\section{Background and Related Works}
\label{sec:background}
\input{7_related}

\section{Security Classification Tasks}
\label{sec:security_tasks}
\input{3_1_security_tasks}

\section{The FL Settings}
\label{sec:fl_settings}
\input{3_2_fl_setting}

\section{The Effectiveness of FL}
\label{sec:performance}
\input{4_performance.tex}
\section{The Byzantine Resilience of FL}
\label{sec:adversary}
\input{5_1_adversarial_robustness.tex}

\section{The Efficiency of FL}
\label{sec:efficiency}
\input{5_2_fl_efficiency}

\section{Discussion}
\label{sec:discuss}
\input{6_discuss.tex}


\section{Concluding Remarks}
\label{sec:conclusion}
\input{conclusion.tex} 

%% file: 2_intro.tex
Machine learning especially deep learning has been increasingly adopted in the cyber security domain, along with good performance achieved in various cyber threat detection tasks, e.g., malware detection~\cite{arp2014drebin, McLaughlin2017}, spam filtering~\cite{roy2020deep, ghourabi2020hybrid, abayomi2022deep,liu2021spam, wu2017twitter}, malicious network traffic detection~\cite{lyamin2018ai, shafiq2020corrauc}.  However, the power of machine learning is hindered by several security-specific critical factors.  First of all, it is becoming more and more challenging for security vendors to collect individual-relevant and privacy-sensitive threat datasets (e.g., SMS spam/non-spam messages, and dynamic runtime logs of Android apps), which are essential for building up robust and up-to-date machine learning models. This is partly due to the enforcement of various regulations on data protection and privacy, such as General Data Protection Regulation (GDPR)~\cite{hoofnagle2019european} and California Consumer Privacy Act (CCPA)~\cite{harding2019understanding}. These regulations aim to enhance individuals’ control and rights over their personal data, along with more limitations enforced for data controllers and processors including security vendors. Then, following these privacy regulations, computing platforms (e.g., Android and iOS) have updated their developer policies along with more restrictions on collecting and uploading on-device data~\cite{android_privacy}. What makes things even worse is that the robustness of cyber threat detection models largely relies on the freshness of the training dataset. This is because of the issue of concept drift~\cite{jordaney2017transcend, xu2019droidevolver}, i.e., the aging of the detection models  as the attackers keep evolving their malicious payloads to escape from the detection radar. Also, the inability of collecting up-to-date threat data has already led to degraded performance for some critical security detection tasks, e.g., SMS spam filtering~\cite{tang2022clues}.

Another factor resides in the curse of isolated data islands. A typical security vendor will usually devote much effort to collecting cyber threat data from various sources, and regular organizations (e.g., governments) can also observe cyber threats during their security operation, such as DDoS attacks targeting their network infrastructure or spam emails targeting their employees. 
In many cases, sharing such threat data across organizations can help build up a  more capable threat detection model. However, the curse of isolated data islands is there, as such threat incidents, especially ones observed during security operations could contain sensitive information relevant to either the involved organizations or individuals. Sharing such threat data will likely incur non-negligible risks for the involved parties (e.g., compromising an organization’s reputation or business interests).

To address these obstacles, we explore in this study the (in)feasibility of federated learning for privacy-preserving cyber threat detection, which, to the best of our knowledge, is among the first-of-its-kind works (see \S\ref{sec:background} for more discussion).  
Federated learning (FL), as a distributed learning paradigm, has been well explored in multiple privacy-sensitive domains, e.g., medical imaging understanding~\cite{sheller2019multi} and the next-word prediction task in virtual keyboards~\cite{hard2018federated}. In a nutshell, FL allows distributed clients (data owners) to collaboratively train a machine learning model without revealing their local datasets to any parties. Despite achieving promising performance results in some privacy-sensitive machine learning tasks, FL is also known to be vulnerable to adversarial machine learning attacks, especially poisoning attacks~\cite{NIPS2017, bhagoji2019analyzing, tolpegin2020data, baruch2019little, sun2019can, bagdasaryan2020backdoor, wang2020attack}. Therefore, this study focuses on the following research questions.  
 First of all, \textit{how effective can FL be when applied to cyber threat detection tasks?} In another word, compared to centrally-trained counterparts, what performance can FL-trained threat detection models achieve, especially in security-specific realistic settings? Then,  compared to non-security prediction models, the FL training process for threat detection models can be subject to more adversarial attacks, i.e., various poisoning attacks.  Therefore, the secondary research question is \textit{how adversary-resistant the FL training process can be when evaluated under a realistic threat model.}  In addition to effectiveness and adversarial resistance, one more research question we aim to understand is that \textit{how (in)efficient the FL training process can be when applied to cyber threat detection tasks}.


To pursue aforementioned research questions, We have encountered and addressed several challenges. Firstly, we need to decide what threat detection tasks to target. A qualified task should satisfy the following three criteria: 1) It should be a well-known and representative threat detection task; 2) It involves privacy-sensitive scenarios, e.g., collecting data from end users; 3) It may gain benefits through adopting FL. Following these criteria,  SMS spam detection and Android malware detection, two binary classification tasks, are chosen, for which, detailed explanations will be given in \S\ref{sec:security_tasks}.
Briefly, the SMS spam detection task requires collection of privacy-sensitive SMS spam/non-spam messages, which can thus benefit from FL. On the other hand, 
many Android malware detection techniques rely on dynamic features that are extracted from runtime logs of Android apps. Such runtime logs are critical but
costly to collect on the server, but are abundant on end-user devices. However, similar to SMS messages, collecting app runtime logs from end-user devices is privacy-sensitive. 
 Then, given the threat detection tasks, one more issue we encountered is that either the datasets or the classification algorithms evaluated in previous works are outdated. For instance, when training and evaluating the proposed SMS spam detection models, most previous studies used only the UCI SMS spam dataset~\cite{almeida2011contributions} which was published more than 10 years ago and contains only hundreds of SMS spam messages. Also, although transformer-based language models have achieved state-of-the-art performance in many NLP tasks~\cite{devlin2018bert, koroteev2021bert}, previous works in SMS spam detection failed to consider such kinds of latest machine learning advancements. On the other hand, most studies in Android malware detection differ in their evaluation datasets, which impedes a direct and objective comparison among these studies.  Therefore, we addressed this issue through collecting up-to-date and large-scaled threat datasets as well as considering the latest progress in machine learning when building up threat detection models. 

Given threat datasets and machine learning algorithms decided, the third challenge stands out as what FL experiments to conduct so as to evaluate FL in a realistic and security-specific manner. To achieve this goal, multiple novel experiments have been designed and implemented, which take security-specific scenarios and realistic threat models into consideration. Particularly, we observe that FL clients in threat detection tasks tend to follow various non-IID data distributions. 
For instance, FL clients in the SMS spam detection task can vary a lot in terms of the natural languages of their local SMS messages, while the Android malware detection task may involve a non-IID distribution in terms of malware families across FL clients. Besides,
different from non-security prediction tasks, FL clients in threat detection tasks may consistently have samples of one class outweigh samples of the other class, e.g., all FL clients have much more benign SMS messages than spam messages. We name this scenario as \textit{the consistent label imbalance (CLI)} scenario. Therefore, when evaluating the effectiveness of FL, we focus on various non-IID  distributions of the threat samples across FL clients, rather than exhaustively exploring FL hyperparameters. Also, when evaluating the adversary resistance of FL, we define a realistic threat model and dedicate our evaluations to practical adversary settings, e.g., the fraction of compromised FL clients $M$ is considered as practical only when $M \leq 5\%$ for data poisoning and $M \leq 1\%$ for model poisoning.  



Next, we highlight some key observations and findings as distilled from our FL experiments.  
First, regarding the effectiveness of FL, we found that cyber threat detection models trained via FL can achieve a performance that is comparable to that of their centrally trained counterparts. For instance, the spam detection model trained via cross-device FL has achieved a recall of 99.03\% and a precision of 99.05\% , while it is 98.79\% and 99.23\% respectively for the centrally trained counterpart.  Then, in realistic FL deployments, there can be various non-IID data distributions across FL clients. And our evaluations show that non-IID data distributions can lead to high instability during the convergence of the FL training process, while the impact on the model performance is either negligible or minor. Particularly, a high degree of quantity-based non-IID distribution can even yield better model performance and faster training convergence, while a high degree of label-based non-IID distribution has no obvious impact on model performance but can incur notable fluctuations (i.e., unstable convergence) in FL training, especially for the cross-device FL. Then, when it comes to  the consistent label imbalance scenario that is specific to cyber threat detection tasks, we observed that CLI towards positive (malicious) samples can incur a non-negligible degradation in performance, while the performance impact of CLI towards negative samples (more practical in cross-device FL) is minor. 

Regarding the adversarial resistance of FL (\S\ref{sec:adversary}), we observed that data poisoning with a practical fraction of poisoned clients ($\leq 5\%$) has a \textit{negligible} attack impact of up to 0.14\% decrease in model accuracy for both the SMS spam detection task and the Android malware detection task. On the other hand, for model poisoning  with a practical fraction of poisoned clients ($\leq 1\%$), the attack impact is also \textit{negligible} for Android malware detection as well as being minor for SMS spam detection with up to 1.52\% decrease in accuracy. Furthermore, the minor performance impact of practical model poisoning can be addressed through the deployment of robust aggregation rules (e.g., Trimmed Mean~\cite{pmlr-v80-yin18a} and Multi-Krum~\cite{NIPS2017}). For instance, Trimmed Mean has decreased the attack impact of model poisoning on SMS spam detection to almost zero (0.09\%). Then, regarding the efficiency of FL, we found out that non-IID data distributions can incur a significant delay in convergence, to address which, a bootstrapping strategy has been proposed with its effectiveness well demonstrated (\S\ref{sec:efficiency}).


Our key contributions are two-fold. On one hand, we have systematically evaluated the effectiveness,  byzantine resilience, and efficiency of FL for two representative cyber threat detection tasks, which features security-specific and realistic FL experiments. On the other hand, our evaluation has distilled a set of novel findings regarding the pros and cons of FL when applied to privacy-sensitive threat detection tasks, which we believe can benefit future efforts of fostering FL-based privacy-preserving cyber threat detection.

%% file: 7_related.tex
\subject{Federated learning.} As an emerging distributed learning paradigm, federated learning (FL)~\cite{McMahan2017} enables distributed FL clients (i.e., mobile devices) to jointly train a global deep learning model without the necessity of sharing their local privacy-sensitive data. 
In recent years, FL has been explored in multiple privacy-sensitive areas~\cite{hard2018federated, brisimi2018federated, sheller2019multi, dayan2021federated, zhang2021federated, Campos2022}, e.g., next-word prediction and medical image segmentation. 
Particularly, Hard et al.~\cite{hard2018federated} applied FL to the training of a next-word prediction model that is used in a mobile virtual keyboard. Upon real-world but privacy-sensitive keyboard input data distributed across mobile devices, the FL-enabled next-word prediction model has achieved better performance than the centrally trained counterpart. Besides, FL has also been demonstrated in the task of semantic segmentation of medical image~\cite{sheller2019multi}, which enables cross-institution model training without sharing the medical images of patients. 

In FL, a central server is deployed to instruct client-side local training, aggregate model updates from clients, and evaluate the resulting global model. A typical FL training process consists of multiple rounds of client-side local training and server-side aggregation. 
For instance, in round $r$, the aggregation server first select $n$ out of all the available $N$ workers, and pushes the latest global model (model parameters, i.e., $\theta^r_g$) to the selected $n$ workers. Along with the global model are some configurations to specify the local training, e.g., the number of epochs, the batch size, and the learning rate, etc. Then, each selected client $c$ fine-tunes $\theta^r_g$ with its local private data using stochastic gradient descent (SGD), and gets the updated model $\theta^r_c$. The difference between the client-received global model $\theta^r_g$ and the updated client-specific model $\theta^r_c$ is calculated as $\nabla^r_c= \theta^r_c-\theta^r_g$, and will be uploaded to the server for aggregation. In the server-side aggregation, given $\{\nabla^r_c, c\in[n]\}$, an aggregation rule $f_{\mathsf{agr}}$ will be applied to get $\nabla^r_\mathsf{agr}$, which will be used to update the global model through $\theta^{r + 1}_{g} = \theta^r_g + \eta\nabla^r_\mathsf{agr}$ with $\eta$ being the server-side learning rate. 

Besides, FL is considered as \textit{cross-device} FL when the clients are resource-constrained and large-scaled, e.g., many thousands of mobile devices. In cross-device FL, the client devices are likely crowd-sourced and can thus be byzantine or even malicious. On the other hand, when FL is applied to the training collaboration among trusted organizations, it is called \textit{cross-silo} FL in which the clients are the participating organizations and thus of a smaller scale (e.g., $\leq 100$). 
Since cross-silo FL tends to be a contracted collaboration across large corporations, it is less likely to have byzantine or malicious clients compared with cross-device FL.

A key element of FL is the aggregation rule (AGR). Among a variety of AGRs~\cite{McMahan2017,Li2018, pmlr-v119-karimireddy20a, dimitriadis2020federated, Wang2020}, the most commonly used one is Federated Averaging (FedAVG)~\cite{McMahan2017}, wherein the server updates the global model through dimension-wise and weighted averaging of the model updates. 
Besides, despite avoiding uploading private client-side data, the uploaded model updates, if leaked to attackers,  may still raise various privacy attacks, e.g., gradient inversion attack~\cite{Hatamizadeh2023, geiping2020inverting,zhu2019deep}, preference profiling attack~\cite{Zhou2022} and membership inference attack~\cite{Nasr2019}. 
To mitigate privacy attacks, various secure aggregation protocols~\cite{bonawitz2017practical, bell2020secure, so2021turbo, guo2022microfedml} have been proposed to ensure the aggregation can be carried out without revealing local model updates.

However, these privacy attacks require the access to model updates of benign FL clients and thus assume that the attackers can either compromise the central server or conduct successful man-in-the-middle (MITM) attacks against the communication between a targeted client and the central server, which we consider as impractical when evaluating the adversarial resistance of FL. Instead, we consider a practical threat model wherein the central FL server is trusted and the communication between FL clients and the central server is free of MITM attacks. Therefore, privacy attacks and secure aggregation rules are not evaluated in this study.

\subject{Poisoning attacks against FL.} 
Depending on the attacker's goal, poisoning attacks against FL can be divided into three subcategories -- \textit{targeted} FL poisoning, \textit{backdoor} FL poisoning and \textit{untargeted} FL poisoning. The \textit{targeted} FL poisoning attacks~\cite{bhagoji2019analyzing, tolpegin2020data} target a subset of designated classes and aim to decrease the model's prediction performance only for the classes of interest. Similarly, the \textit{backdoor} FL poisoning attacks~\cite{baruch2019little, sun2019can, bagdasaryan2020backdoor, wang2020attack} are designed to ensure that the corrupted model predict normally for regular samples but fails for samples stamped with a backdoor pattern (e.g., special pixel values in an image), and can thus be considered as a variant of the targeted FL poisoning attack. On the contrary, 
an \textit{untargeted} FL poisoning attack~\cite{NIPS2017, baruch2019little, Fang2019, shejwalkar2021manipulating} is intended to degrade the overall prediction accuracy of a model for any input across all the classes. When evaluating adversary resistance of FL for security tasks, we focus on untargeted FL poisoning attacks, as they are more challenging to carry out and can incur great threats to real-world FL deployments~\cite{shejwalkar2022back}.

Similar to poisoning attacks against central training, poisoning attacks against FL can be carried out through compromising the local datasets (e.g., label flipping), namely, \textit{data poisoning} attacks~\cite{wang2020attack, tolpegin2020data, shejwalkar2022back}.  Furthermore, 
attackers may also directly manipulate the model updates of compromised devices, and thus poison the resulting model, which is named as the \textit{model poisoning} attack~\cite{NIPS2017, bhagoji2019analyzing, sun2019can,bagdasaryan2020backdoor, wang2020attack, Fang2019, shejwalkar2021manipulating, shejwalkar2022back}. 
Existing model poisoning attacks against FL aim to directly manipulate the model updates of compromised devices in a manner that maximizes the attack goal while satisfying various constraints with regards to the fraction of compromised clients and the norms of the manipulated model updates. 
Representative model poisoning attacks include the \textit{Little is Enough} (LIE) attack~\cite{baruch2019little}, MIN-MAX and MIN-SUM~\cite{shejwalkar2021manipulating}, static optimization (STAT-OPT)~\cite{Fang2019} , and projected gradient ascent (PGA)~\cite{shejwalkar2022back}. In this study, we assume a practical threat model featuring that the attacker is AGR-agnostic, therefore, three AGR-agnostic model poisoning algorithms are considered, which include LIE, MIN-MAX, and MIN-SUM.

\subject{Defense against FL poisoning attacks.} To mitigate poisoning attacks, various robust aggregation rules (AGRs) have been proposed and evaluated. Such robust AGRs can be classified into two groups depending on how they detect and minimize the impact of manipulated model updates (gradients). One group focuses on filtering out malicious gradients either dimension-wise or vector-wise, and concrete examples include Krum~\cite{NIPS2017}, Multi-Krum~\cite{NIPS2017}, median-based filtering~\cite{chen2017distributed, pmlr-v80-yin18a, xie2018generalized, pillutla2022robust}, Bulyan~\cite{guerraoui2018hidden}, ERR/LFR~\cite{Fang2019}, divide-and-conquer (DnC)~\cite{shejwalkar2021manipulating}.  The other group of AGRs utilizes various regularization techniques to minimize the adversarial impact of malicious model updates, e.g., Trimmed Mean~\cite{pmlr-v80-yin18a}, norm clipping~\cite{sun2019can}. In this study, when evaluating the effect of robust AGRs, we select from each group a representative robust AGR, namely, Multi-Krum and Trimmed Mean respectively.

\subject{Machine learning based SMS spam detection.} 
Spam denotes unsolicited messages that are delivered to victims through various channels, such as the short message service (SMS) and the Email services. Varied by the delivery channel and the message format, spam messages can thus be grouped as SMS spam, Email spam, among others. In this study, we focus on the  SMS spam detection considering the following factors. Particularly, it is privacy-invasive to collect SMS messages, regardless of spam or not spam, not to mention uploading SMS messages to the central server for model training and evaluation, which partially explains the scarcity of publicly available SMS spam/non-spam datasets. Besides, SMS messages are short in their length, and the underlying SMS spam campaigns keep evolving across time with more stealthy evasion techniques~\cite{tang2022clues}, rendering existing detection systems decay in their performance. All these factors make SMS spam detection a perfect scenario for federated learning. 

The detection of SMS spam is typically defined as a binary text classification task. To conquer this task, various machine learning algorithms have been explored, which range from traditional classification algorithms (e.g., naive Bayes, random forest, and support vector machine )~\cite{almeida2011contributions, xu2012sms, sjarif2019sms}, to deep neural network architectures such as CNN~\cite{roy2020deep, ghourabi2020hybrid}, LSTM~\cite{roy2020deep, ghourabi2020hybrid}, BiLSTM~\cite{abayomi2022deep}, and transformer models~\cite{liu2021spam}. 
Particularly, Roy et al.~\cite{roy2020deep} built up a CNN-based SMS spam detection model which has achieved a state-of-the-art performance when trained and evaluated on the UCI SMS spam dataset~\cite{almeida2011contributions}. However, almost all these detection models~\cite{almeida2011contributions, sjarif2019sms, roy2020deep, liu2021spam, abayomi2022deep} were built upon the UCI SMS spam dataset~\cite{almeida2011contributions} which, released in 2012, is small-scaled with only 747 English spam messages and likely outdated. To address the scarcity of public SMS spam datasets, Tang et al.~\cite{tang2022clues} proposed \textit{SpamHunter} to collect SMS spam messages as reported by victims on Twitter, which results in the release of 22K multilingual SMS spam messages, while Abayomi-Alli et al.~\cite{abayomi2022deep} released another English spam dataset coined as \textit{ExAIS\_SMS} which was collected from 20 end users. In this study, we have combined all these newly released datasets when building up and evaluating SMS spam detection models in FL settings.

\subject{Machine learning based Android malware detection.}
Malware detection aims to decide whether a given software program is malicious or not, and optionally attribute a malicious program (malware) to relevant malware families. Serving as a key threat detection task, malware detection plays an important role in protecting end users and end devices. 
As malware programs on different operating systems can have significantly differentiated syntactic and malicious behaviors, the malware detection problem can be further divided into Android malware detection, Windows malware detection, and iOS malware detection, among others. 
In this study, we choose Android malware detection as the 2nd threat detection task due to several factors elaborated below. Firstly, Android, as an open mobile operating system, has been widely adopted by billions of mobile users, which gives rise to a large client base for FL-enforced malware detection. 
Also, alike SMS spam detection, Android malware detection also suffers from the scarcity and outdatedness of publicly available datasets~\cite{ rahali2020didroid, wang2022malradar}. Besides, as malware intelligence is an important intellectual property, it is impractical for security vendors to directly share their proprietary malware datasets. Instead, collaborative learning via FL appears to be promising, which will be comprehensively evaluated in this study. 

A long line of studies have thus proposed various machine learning systems for Android malware detection~\cite{arp2014drebin, mariconti2016mamadroid,McLaughlin2017, karbab2018maldozer, wang2019effective, xu2019droidevolver, rahali2020didroid, gao2021gdroid, hei2021hawk}, e.g., Drebin~\cite{arp2014drebin}, Mamadroid~\cite{mariconti2016mamadroid}, and MalDozer~\cite{karbab2018maldozer}. 
These systems differ in many aspects, especially how detection features are extracted, what machine learning algorithms have been adopted, as well as how the detection system is evaluated and what groundtruth datasets have been applied to the evaluation. 
Particularly, McLaughlin et al.~\cite{McLaughlin2017} abstracted an Android app as a sequence of opcodes and utilized an embedding layer to automatically embed the opcode sequence into a fixed-size feature vector, while Gao et al.~\cite{gao2021gdroid} considered an Android app as a node in a graph and utilizes node embedding techniques to auto-encode an Android app. Also, various classification algorithms have been explored, e.g., random forest~\cite{mariconti2016mamadroid}, SVM~\cite{arp2014drebin, nix2017classification}, CNN~\cite{McLaughlin2017, nix2017classification, karbab2018maldozer, wang2019effective}, LSTM~\cite{McLaughlin2017, nix2017classification}, and graph neural networks~\cite{gao2021gdroid, hei2021hawk}. Among these algorithms, CNN has achieved the best performance in multiple evaluations~\cite{McLaughlin2017,nix2017classification}, and we thus choose the CNN architecture proposed in \cite{McLaughlin2017} as the default Android malware detection model for our FL experiments. 

\subject{The evaluation of FL on cyber threat detection tasks.} Concurrent with our work, Sidhpura et al.~\cite{Sidhpura2023} evaluated the effectiveness of FL on SMS spam detection. However, their FL experiment settings are not practical as only two FL clients were deployed and the spam dataset under evaluation was outdated. Another study on FL-based SMS spam detection only considered three clients~\cite{SrinivasaRao2023}. Also, the authors failed to explore security-specific FL scenarios (e.g.,  consistent label imbalance settings), not to mention profiling the adversarial resistance of FL for SMS spam detection. 
Besides, federated SVM was applied in Android malware detection in 2020~\cite{Hsu2020} while a FL-based Android malware detection framework, namely, FEDriod, was introduced in 2023~\cite{Fang2023}. However, both studies considered no more than 7 FL clients for training, emphasizing only the effectiveness rather than efficiency and adversary resistance of federated learning. In 2023, a dynamic weighted federated averaging (DW-FedAvg) strategy was applied to Android malware detection~\cite{Chaudhuri2023}. Besides, Fereidooni, et al.~\cite{fereidooni2022fedcri} applied the concept of cross-silo FL to threat intelligence sharing across organizations. Still, very few settings were considered when profiling the effectiveness of FL, and no experiments were conducted to understand the adversarial resistance of FL.

Moving forward from these works, when evaluating the effectiveness of FL for cyber threat detection tasks, we have comprehensively considered various non-IID settings and consistent label imbalance scenarios. Furthermore, we have profiled, for the first time, the adversarial resistance of FL, in practical threat settings and for cyber threat detection tasks. Lastly, we have highlighted a set of efficiency issues for FL-based cyber threat detection along with promising solutions proposed and demonstrated.

%% file: 3_1_security_tasks.tex
In this study, we target two representative security classification tasks, namely, SMS spam detection and Android malware detection, for both of which, respective detection models have been reproduced with state of the art (SOTA) performance achieved. Below, we elaborate these detection models, their training/testing datasets, as well as the model performance.

\subject{SMS spam detection.} 
As aforementioned (\S\ref{sec:background}), we focus on the \textit{SMS} spam detection because this task is both challenging and privacy-sensitive when compared to other spam detection tasks, e.g., Twitter spam and Email spam. 

\begin{table}
    \centering
    \footnotesize
    \caption{Our SMS Spam datasets. }
    \label{tab:sms_spam_datasets}
    \begin{tabular}{ccccc}
        \toprule
        Datasets & Spam & Non-Spam & Languages & Period\\
         \midrule
        UCI &747 &4,827&English&2012\\
        ExAIS &2,350&2,890&English&2015\\
        SpamHunter & 23,249& 0 & Multilingual & 2018-2022\\
        Twitter & 0& 18,629 & Multilingual & 2018-2022\\
        Total &26,346& 26,346& Multilingual & 2012-2022 \\
         \bottomrule
    \end{tabular}
\end{table}

\subsubject{Datasets.} As listed in Table~\ref{tab:sms_spam_datasets}, our groundtruth dataset is a combination of multiple publicly available SMS spam datasets. The first is the UCI SMS spam dataset~\cite{almeida2011contributions} which was released in 2012 and consists of only 747 SMS spam messages and 4,827 non-spam messages. To enable an up-to-date evaluation of SMS spam detection,  two more recent datasets were further collected, namely,  ExAIS~\cite{abayomi2022deep} and SpamHunter~\cite{tang2022clues}. In total, these three datasets have contributed 26,346 SMS spam messages and 7,717 SMS non-spam messages which are featured by 75 different natural languages and over 12 diverse spam categories~\cite{tang2022clues}. Furthermore, to facilitate solid performance evaluation, we moved to make this dataset balanced. Specifically,  we randomly sampled tweets from the Twitter archive~\footnote{https://archive.org/details/twitterarchive} to complement the non-spam SMS messages, which is based upon the assumption that most posts published on Twitter are benign. This assumption is aligned with observations in previous studies~\cite{INUWADUTSE2018496,Varol2017}. Also, we manually labeled 1,000 sampled tweets and confirmed that 97\% of them are benign. In total, we have got a balanced dataset of 52,692 spam/non-spam messages. And we name this dataset as \textit{Spam-2022}.

\subsubject{The model architectures.} Based on a comprehensive literature for spam detection ~\cite{almeida2011contributions, xu2012sms, sjarif2019sms, roy2020deep, ghourabi2020hybrid, abayomi2022deep, liu2021spam}, two representative neural network architectures are selected for SMS spam detection. One is the CNN model as proposed in \cite{roy2020deep}, while the other is a pre-trained transformer-based language model, namely, the multilingual BERT~\cite{devlin2018bert} which has achieved SOTA performance in many NLP tasks~~\cite{devlin2018bert, koroteev2021bert}. Note that the original CNN model supports only English text as input and SMS messages in other languages should be translated into English in advance. Also, the CNN model has only 7 million parameters, which is much smaller than the multilingual BERT model of 167 million parameters. 

\subsubject{The model performance.} Given the largest-ever SMS spam groundtruth dataset (Spam-2022), two SMS spam detection models have been trained and tuned with different hyper-parameters explored.
During model training,  80\% of the spam groundtruth were randomly sampled out for training and validation,  while the left 20\% were used for testing.
Table~\ref{tab:spam_detection_models} lists a direct comparison between the CNN model and the BERT model with regards to their spam detection performance. As we can see, the BERT model outperforms the CNN model by 1.76\% in recall and by 0.90\% in F1 score. Considering the better performance as well as the multilingual support, the BERT model was chosen as the baseline model in our FL experiments for SMS spam detection.   


\begin{table}
    \centering
    \footnotesize
    \caption{SMS Spam detection models.}
    \label{tab:spam_detection_models}
    \begin{threeparttable}
    \begin{tabular}{ccccc}
        \toprule
        Model & Datasets & Recall & Precision & F1-Score \\
         \midrule
        CNN~\cite{roy2020deep}& Spam-2022~\tnote{1} & 0.9788 & 0.9807 & 0.9797 \\
         BERT~\cite{devlin2018bert}& Spam-2022  & 0.9879&0.9923&0.9804\\
         \bottomrule
    \end{tabular}
    \begin{tablenotes}
        \item [1] This is a merge of all datasets listed in Table~\ref{tab:sms_spam_datasets}.
    \end{tablenotes}
    \end{threeparttable}
\end{table}

\begin{table}
    \centering
    \footnotesize
    \caption{Our Android malware datasets. }
    \label{tab:malware_datasets}
    \begin{tabular}{ccccc}
        \toprule
        Datasets & Malware & Benign & Period\\
         \midrule
        Drebin & 4255 & 0 &2010-2012\\
        CIC-AndMal2017 & 0 &1645&2015-2017\\
        Androzoo & 0 & 2610 & 2022\\
        Total &4255& 4255& 2010-2022 \\
         \bottomrule
    \end{tabular}
\end{table}

\subject{Android malware detection.} For Android malware detection, cross-silo FL may help security vendors collaboratively train an effective Android malware detection without revealing their proprietary malware datasets, while cross-device FL provides a possibility for Android users to contribute to Android malware detection without the need of uploading their local malware/benign Android apps to the server. Although benign Android apps can be easily collected, it is not the case for their realistic runtime logs, e.g., the api call sequence and network traffic flows. Such dynamic features are critical but costly to collect. However, end-user devices have app logs that are both realistic and abundant, but also privacy-sensitive,  which renders a good fit for cross-device FL. 

\subsubject{Datasets.} Following previous practices in Andriod malware detection, a balanced groundtruth dataset has been composed, which consists of 4255 malware samples and 4255 benign samples as listed in Table~\ref{tab:malware_datasets}.
For the malware samples, all were collected from the Drebin dataset~\cite{arp2014drebin}, one of the most commonly used malware dataset~\cite{Campos2022,karbab2018maldozer,277204Dos}. 
Specifically, the Drebin dataset contains 5560 malware samples that belong to 179 different malware categories and families. After excluding files with decompiling failures or files with size smaller than 5KB, 4255 were selected out as the malware samples in our groundtruth. 
Since Drebin doesn't contain benign Android apps, the benign samples were composed from two sources. One is CIC-AndMal2017~\cite{lashkari2018toward} and the other is Androzoo~\cite{AndroZoo}. Benign samples in the CIC-AndMal2017 were apps published in Google Play between 2015 and 2017. To guarantee the variety and obtain new samples, we also downloaded some other benign samples published in 2020 from Androzoo.
For all benign samples, we queried VirusTotal~\cite{virustotal} regarding their maliciousness and have thus confirmed there are no virus alerts for each of the selected benign samples. 

\subsubject{The model architecture.} We choose the CNN architecture proposed in \cite{McLaughlin2017} as the baseline model for Android malware detection, owing to its multiple advantages over other works~\cite{arp2014drebin, mariconti2016mamadroid, karbab2018maldozer, wang2019effective, xu2019droidevolver, rahali2020didroid, gao2021gdroid, hei2021hawk}. First of all, it has achieved a SOTA performance in multiple evaluations.
Then, it takes as input only static features of an Android app, specifically, the sequence of opcodes, which can be easily extracted with low computing overhead. 
On contrary, other Android malware detection methodologies extract features either through dynamic execution of a given app~\cite{Zhang2020, mariconti2016mamadroid} or complicated static analysis~\cite{Jindal2019}, both of which are costly and thus infeasible in the FL scenario especially for cross-device FL. 
What's more, \cite{McLaughlin2017} is among the very few works that have released the source code for training and evaluating the proposed malware detection systems, which makes it time-efficient for us to reproduce their model.

\subsubject{The model performance.} Given the model architecture and datasets, a baseline model was then trained on our central server so as to facilitate a direct comparison with future models trained through federated learning. When building up this baseline model, 90\% ground truth was used for training, and 10\% were held out for model testing. 
Then, when training the model, the  learning rate was set up as $1e^{-4}$ while the batch size was 5. As a result, this baseline model has achieved a precision of 98.60\%, a recall of 99.76\%, an accuracy of 99.18\%, and a F1-score of 99.18\%.

%% file: 3_2_fl_setting.tex
In this study, we aim to profile the feasibility of FL for the aforementioned two security classification tasks, namely SMS spam detection, and Android malware detection. This is achieved from three perspectives: effectiveness, byzantine resilience (i.e., adversarial resistance), and efficiency. Instead of exhaustively evaluating all possible FL configurations (hyper-parameters), we focus on \textit{realistic} and \textit{security-specific} scenarios which were missed by most previous works but are critical for the real-world FL deployment for cyber threat detection.
Below, we first present the default FL settings (e.g., FL framework, hyper-parameters, and the computing environment) that will be adopted across our FL experiments, if not otherwise specified.
What is followed is  an overview regarding what experiments we have designed and conducted so as to give you the whole picture before going into detailed experimental results, which will be presented in \S\ref{sec:performance} for the effectiveness of FL, \S\ref{sec:adversary} for the byzantine resilience, and \S\ref{sec:efficiency} for the efficiency (cost) of FL.

\subject{An overview of the FL experiments.} Our FL experiments are designed to profile the following questions. First of all, \textit{How effective can FL be when applied to representative security prediction tasks in realistic settings?} Then, given byzantine or even malicious clients that are likely to be present in security prediction scenarios, \textit{how byzantine-resilient can FL be?} Besides, with or without defenses against adversarial attacks, \textit{how efficient can FL be}? Finally, given FL evaluated in terms of effectiveness, byzantine resilience, and efficiency, \textit{how can we address the weaknesses of FL if any?} In other words, what are the best practices when applying FL to security prediction tasks?

\subsubject{Experiments to evaluate FL effectiveness.} The first set of FL experiments are designed to evaluate the effectiveness of FL in realistic settings for security prediction tasks. Instead of exhausting FL hyper-parameters,   experiments in this group focus on evaluating realistic categories of non-IID (non-independent and identically distributed) data distributions across FL clients. 
Two of them are the non-IID data distributions of samples in terms of labels or quantity, which have also been evaluated in previous FL works for non-security tasks~\cite{Li2022}, and we thus name them as label-based non-IID and quantity-based non-IID. 
Taking the label-based non-IID for SMS spam detection as an example, one client in the cross-device FL may have a spam/non-spam $\alpha$ ratio of $2:1$ while another may have an inverse $\alpha$ ratio of $1:2$. 

Then, a security-specific non-IID scenario is also designed and evaluated. Specifically, in our cross-device FL for spam detection, clients (i.e., real-world end devices, especially mobile devices) tend to consistently have more non-spam samples than spam ones in their local datasets, whereas organization participants in the cross-silo FL are likely to have lots of spam messages but much fewer non-spam ones due to the privacy regulations. This gives rise to a scenario that is different from traditional non-IID data distributions, and we call it \textit{the consistent label imbalance} (CLI) wherein most (if not all) clients consistently have samples of one class outweighing samples of another class. 
This CLI scenario is also applicable to malware detection, especially for the cross-device FL wherein the participating Android end devices are likely to have more benign apps installed than the malware ones. 
However, this may not be the case for the cross-silo malware detection as the participating anti-virus vendors are likely to have balanced datasets.

Besides, in the spam detection task, FL clients located in different countries or regions may differ in the language of their local message datasets. We name such kinds of scenarios as \textit{language-based non-IID}. Similarly, Android malware apps may belong to different malware families, and samples belonging to different malware families can differ in their non-IID distribution across clients, which will also be evaluated under the name of \textit{family-based non-IID}.
More detailed experiment settings are presented in Section~\ref{sec:performance} along with respective results and observations. 

\subsubject{Experiments to evaluate FL byzantine resilience.} Since the FL clients can be byzantine or even malicious especially in the cross-device FL, the 2nd group of experiments aim to evaluate the byzantine resilience (adversary resistance) of FL in security prediction tasks. To achieve this, we first evaluated a representative data poisoning attack, namely label flipping~\cite{shejwalkar2022back}, under different ratios of compromised clients. Also, previous works typically assume the attacker can poison all samples on a client or even generate a poisoned dataset that is much larger than those of benign clients, however, we argue that this is not realistic. This is because most attackers typically have no direct control over a large volume of FL clients, but can only remotely distribute poisoned data samples to FL clients, which means they can only poison a portion instead of all of the local samples. For instance, in the SMS spam detection scenario, attackers may distribute stealthy but true spam messages to some FL clients, and these messages may escape the device owner's annotation and be considered as non-spam during FL training. However, in the meantime, the attacker cannot prevent the same FL client from receiving SMS messages (spam or non-spam) from other sources as well as assigning true labels to these messages. The only chance that an attacker can fully poison a client is when they have full control of the respective client, in which case  model poisoning is a better option rather than data poisoning. Also it can be easily detected by the server if an unusually large set of poisoned samples is used by the attacker.  Therefore, stepping forward from previous studies, we evaluated different poisoning rates, i.e., the fraction of local samples being poisoned, and we assume that the size of the poisoned dataset is similar to that of a regular dataset of a benign FL client. 

Then, when some participating FL clients are fully controlled by the attackers, model poisoning attacks become possible, in addition to data poisoning attacks.  Among the SOTA model poisoning attacks, some (e.g., STAT-OPT and DYN-OPT~\cite{shejwalkar2022back}) rely on the knowledge of the server-side AGR algorithm, which we believe is not practical in the real threat model. Among the left ones without such a dependency, three are selected for our experiments, which include LIE~\cite{baruch2019little}, MIN-SUM~\cite{shejwalkar2021manipulating}, and MIN-MAX~\cite{shejwalkar2021manipulating}.

Note that for the above data/model poisoning attacks, we focus on untargeted poisoning, i.e., the attacking goal is to undermine the availability of the resulting model. We choose to evaluate untargeted poisoning rather than targeted or backdoor poisoning  because untargeted poisoning is more challenging and has higher adversarial impact on realistic FL security applications. We have also evaluated the defensive effectiveness of two representative robust AGRs including Trimmed Mean~\cite{pmlr-v80-yin18a} and Multi-Krum~\cite{NIPS2017}. More details can be found in Section \ref{sec:adversary}.

\subsubject{Experiments to evaluate FL efficiency and identify best FL practices.} We have also characterized FL efficiency through a combination of empirical evaluation and theoretical analysis. Particularly, we investigated how to speed up FL training using model pre-training, especially in scenarios where data imbalance issues significantly slow down model convergence. The full details will be presented in Section \ref{sec:efficiency}.

\begin{table}
    \centering
    \footnotesize
    \caption{The default FL parameters.}
    \label{tab:default_fl_setting}
    \begin{tabular}{
    m{0.45\linewidth}
    >{\centering\arraybackslash}m{0.2\linewidth}
    >{\centering\arraybackslash}m{0.2\linewidth}
    }
        \toprule
        Parameter &  Cross-Device & Cross-Silo\\
        \midrule
         $N$: total number of FL clients& 200 & 20\\
         \hline
          $n$: number of FL clients sampled in each FL round& 10\% of 
          $N$ (20) & All\\
                 \hline
          Aggregation rule (AGR) & \multicolumn{2}{c}{FedAVG}\\
          \hline
          Local epoch number $e$ & \multicolumn{2}{c}{2}\\
          \hline
          \multirow{2}{1.0\linewidth}{Local batch size $\beta$ and learning rate $e$}  & \multicolumn{2}{c}{32, $5e^{-5}$ for SMS spam}\\ 
            & \multicolumn{2}{c}{5, $1e^{-4}$ for Android malware}\\
          \hline
          Data distribution across clients & \multicolumn{2}{c}{Dirichlet distribution with $\alpha = 1$}\\
          \hline
          The split ratio of training and testing & \multicolumn{2}{c}{9:1}\\
        \bottomrule
    \end{tabular}
\end{table}

\subject{The FL framework.} We explored FL frameworks that are both popular and open source, and chose the Flower framework~\cite{beutel2020flower} because of its advantages in many aspects. Particularly, it is extensible enough to allow us to customize the FL training process and experiment with different realistic settings such as the presence of model poisoning attacks and the deployment of robust aggregation rules (AGRs). Besides, it is agnostic to the underlying machine learning frameworks, which facilitates the migration of centrally trained security prediction models to the FL scenario. 

\subject{The default FL hyper-parameters.} Since this study focuses on evaluating realistic and security-specific FL settings, we adopt the same set of FL hyper parameters across experiments, unless specifically noted. Some of these FL hyper parameters are learned from  common practices of previous FL works~\cite{Bonawitz2019, shejwalkar2022back}, e.g., the number of clients in cross-device FL and cross-silo FL, and the aggregation rule, while others are distilled from our preliminary FL experiments, e.g., the client-side learning rate, the number of client-side epochs, etc. 
Also, FL training with these default hyper-parameters has achieved a performance comparable to that of centralized training, for both security classification tasks under our study. One thing to note is that the best hyper parameter setting varies across security classification tasks, and identifying such a setting is not our focus, but can be easily achieved by the central FL server in real-world deployments.

As listed in Table~\ref{tab:default_fl_setting}, the cross-device FL experiments involve the total number of clients $N = 200$ for both security classification tasks. Also, the number of clients sampled in each FL round is configured as $n = 20$. In addition, the well-adopted FedAVG~\cite{McMahan2017} is chosen as the default aggregation rule (AGR) while the maximum of rounds of the FL training is configured as 500. 
Then, the client-side FL settings consist of the local epochs $e = 2$, a learning rate $l = 5e^{-5}$ for SMS spam detection while $1e^{-4}$ for Android malware detection, and a batch size $\beta = 32$ and $5$ similarly.
The cross-silo settings differ in two aspects. Since the cross-silo FL is designed for cross-organization collaborative learning, the total number of clients should be much smaller than that of cross-device FL. Also, unlike the cross-device FL, the cross-silo FL considers all clients (e.g., security vendors) to be available in each round. Therefore, we set $n = N = 20$.
Also, for both cross-device FL and cross-silo FL, 10\% of the samples are held out on the FL server for testing, while the remaining 90\% are distributed to the $N$ clients using a Dirichlet distribution~\cite{dirichlet} with $\alpha = 1$, which gives a quantity-based non-IID distribution of the samples across the FL clients.

\subject{The computing environment.} 
Four servers were set up for our FL experiments.
Each server runs on Ubuntu 20.04.6 LTS and is equipped with an AMD EPYC 7V13 CPU with 24 CPU cores, 220 GB memory, 1 TB of disk storage, and an NVIDIA A100 GPU with 80 GB of GPU memory.

%% file: 4_performance.tex
\begin{table}
    \centering
    \footnotesize
    \caption{The performance of FL-trained baseline models.}
    \label{tab:fl_baseline_model_performance}
    \begin{tabular}{ccccc}
        \toprule
       Task & Model & Precision & Recall & Accuracy  \\
       \midrule
       \multirow{3}{0.1\linewidth}{SMS Spam}  
       & Central & \textbf{0.9923}&$0.9879$&$0.9901$\\
       & $\text{FL}_\text{cross-device}$ & $0.9871$ &$0.9845$ &$0.9858$ \\
       & $\text{FL}_\text{cross-silo}$ & $0.9905$ &\textbf{0.9903} &\textbf{0.9904}\\
       \midrule
        \multirow{3}{0.1\linewidth}{Android Malware}  
         & Central & \textbf{0.9907} & $0.9976$ & \textbf{0.9941} \\
       & $\text{FL}_\text{cross-device}$ & $0.9766$ & $0.9936$ & $0.9849$\\
       & $\text{FL}_\text{cross-silo}$ & $0.9838$ & \textbf{1.0} & $0.9918$\\
       \bottomrule
    \end{tabular}
\end{table}

In this section, we evaluate the effectiveness of FL for security prediction tasks. Below, detailed experiments along with key results and observations are presented.

\finding{The FL-trained security detection models can achieve a performance that is comparable to  their centrally trained counterparts.}

\subject{FL baseline performance.} Using the default FL settings (Table~\ref{tab:default_fl_setting}), a FL baseline model is trained for the aforementioned two security classification tasks and two FL scenarios (cross-device and cross-silo). Table~\ref{tab:fl_baseline_model_performance} lists the resulting performance statistics of these baseline models along with a comparison to centrally trained counterparts. 
We can see that the FL baseline models, no matter whether cross-device FL or cross-silo FL, have achieved a performance that is comparable to their centrally trained counterparts. For instance, the spam detection model trained via cross-device FL has achieved a recall of 99.03\% and a precision of 99.05\% , while it is 98.79\% and 99.23\% respectively for the centrally trained counterpart. 
\begin{figure}
    \centering
    \subfigure[SMS spam classification.]{
        \label{fig:fl_baseline_spam}
        \includegraphics[width=.45\columnwidth]{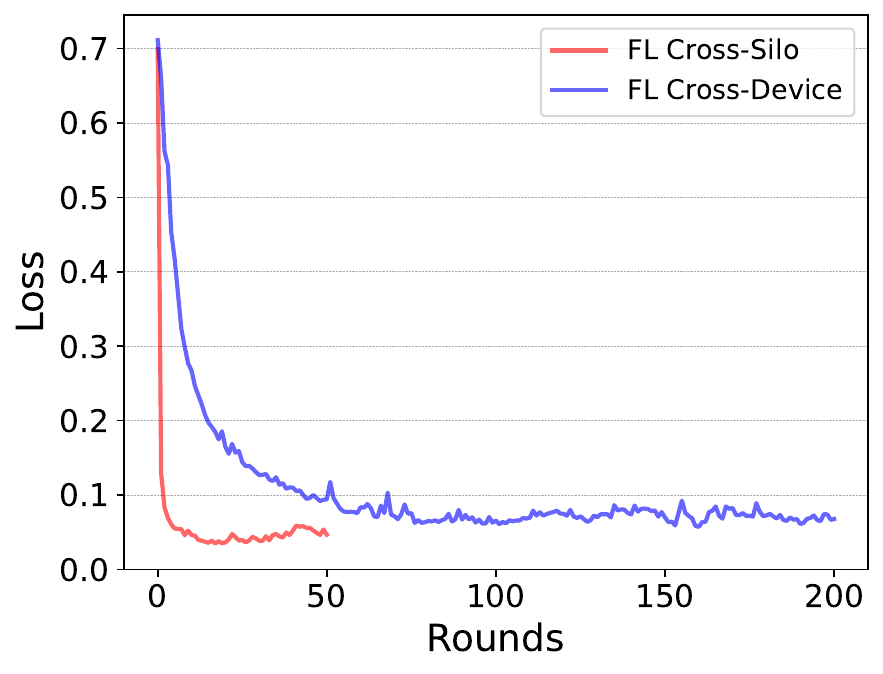}
    }
    \subfigure[Android malware classification.]{
        \label{fig:fl_baseline_malware}
        \includegraphics[width=.45\columnwidth]{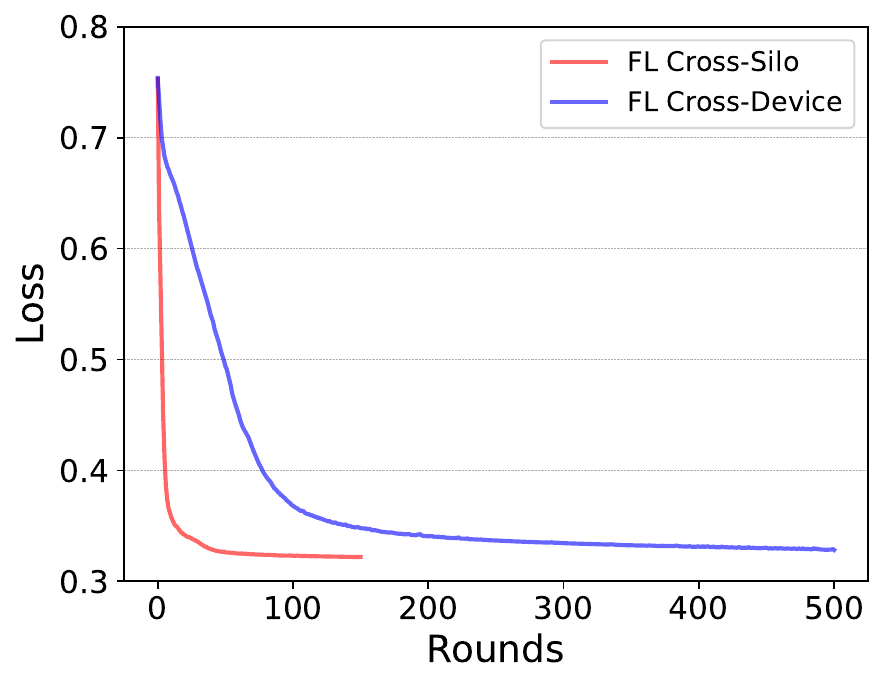}
    }
    \caption{The convergence process of the FL baseline models for SMS spam detection and Android malware classification respectively.}
    \label{fig:fl_baselines}
\end{figure}
Figure~\ref{fig:fl_baselines} also presents the FL training process for these baseline models, and we can see that the cross-silo FL converges faster than the cross-device FL in terms of the number of rounds, which is expected since cross-silo FL involves all clients in each training round while cross-device FL samples only 10\% clients in each training round. In addition, the number of rounds needed to converge varies across security detection tasks. For instance, in cross-device FL training, the SMS spam detection task converges more quickly with less than 100 rounds while it is over 200 rounds for Android malware detection.

\begin{table}
    \centering
    \footnotesize
    \caption{The performance of FL-trained \textbf{spam detection} models under various quantity-based non-IID data distributions.}
    \label{tab:quantity_based_non_iid_spam}
    \begin{threeparttable}
        \begin{tabular}{cccc}
            \toprule
           FL & Dirichlet& Precision & Recall \\
           \midrule
           \multirow{4}{*}{Cross-Device} 
           & $\alpha = 0.5$ & $\mathbf{0.9893}\pm0.0037$ &$\mathbf{0.9865}\pm0.0024$ \\
           &$\alpha = 1$ & $0.9871\pm0.0042$& $0.9845\pm0.0036$ \\
           &$\alpha = 5$ & $0.9806\pm0.0063$& $0.9755\pm0.0037$ \\
           &$\alpha = 10$ & $0.9807\pm0.0062$ &$0.9682\pm0.0046$ \\
           \midrule
            \multirow{4}{*}{Cross-Silo} 
           & $\alpha = 0.5$ & $0.9899\pm0.0015$ & $\mathbf{0.9912}\pm0.0010$\\
           &$\alpha = 1$ &$0.9905\pm0.0025$ &$0.9903\pm0.0027$ \\
           &$\alpha = 5$ &$0.9907\pm0.0019$ &$0.9885\pm0.0019$\\
           &$\alpha = 10$ & $\mathbf{0.9914}\pm0.0015$ & $0.9883\pm0.0017$\\
           \bottomrule
        \end{tabular}
    \end{threeparttable}
\end{table}

\begin{table}[t]
    \centering
    \footnotesize
    \caption{The performance of FL-trained \textbf{malware detection} models under various quantity-based non-IID data distribution.}
    \label{tab:quantity_based_non_iid_malware}
    \begin{threeparttable}
        \begin{tabular}{cccc}
            \toprule
           FL & Dirichlet& Precision & Recall \\
           \midrule
           \multirow{4}{*}{Cross-Device} 
           & $\alpha = 0.5$ & $\mathbf{0.9797}\pm0.0022$ & $\mathbf{0.9976}\pm0.0000$ \\
           &$\alpha = 1$ & $0.9766\pm0.0007$ & $0.9936\pm0.0011$ \\
           &$\alpha = 5$ & $0.9588\pm0.0009$ & $0.9972\pm0.0009$ \\
           &$\alpha = 10$ & $0.9538\pm0.0028$ & $0.9953\pm0.0000$ \\
           \midrule
            \multirow{4}{*}{Cross-Silo} 
           & $\alpha = 0.5$ & $\mathbf{0.9860}\pm0.0000$ & $0.9967\pm0.0012$ \\
           &$\alpha = 1$ & $0.9838\pm0.0000$ & $\mathbf{1.0000}\pm0.0000$ \\
           &$\alpha = 5$ & $0.9751\pm0.0032$ & $0.9939\pm0.0012$ \\
           &$\alpha = 10$ & $0.9633\pm0.0032$ & $0.9953\pm0.0000$ \\
           \bottomrule
        \end{tabular}
    \end{threeparttable}
\end{table}

\begin{figure}
    \centering
    \subfigure[Accuracy.]{
        \label{fl_quantity_non-iid_accuracy}
        \includegraphics[width=.45\columnwidth]{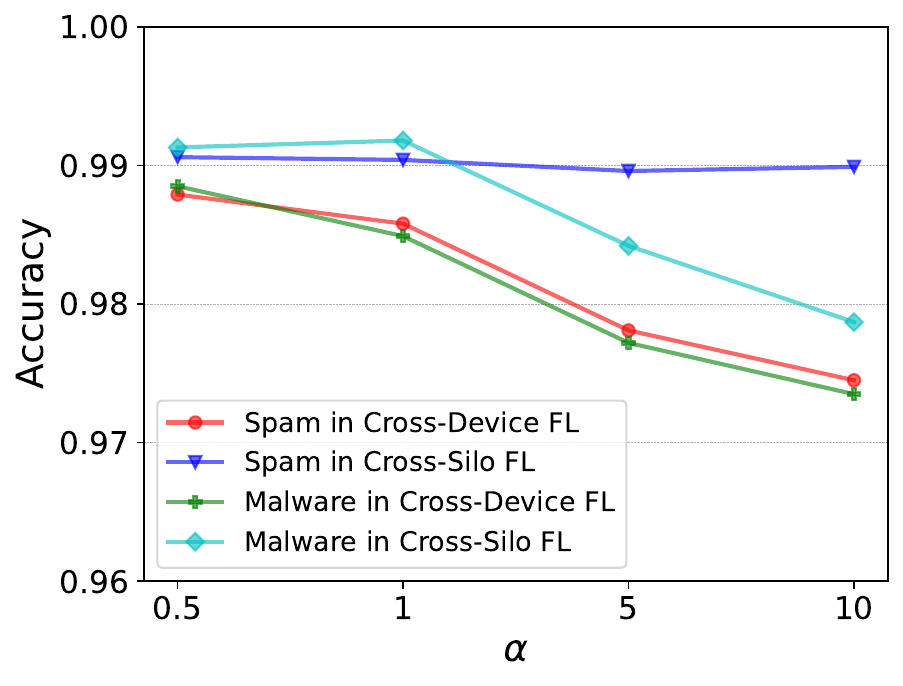}
    }
    \subfigure[The number of rounds for convergence.]{
        \label{fl_quantity_non-iid_convergence}
        \includegraphics[width=.45\columnwidth]{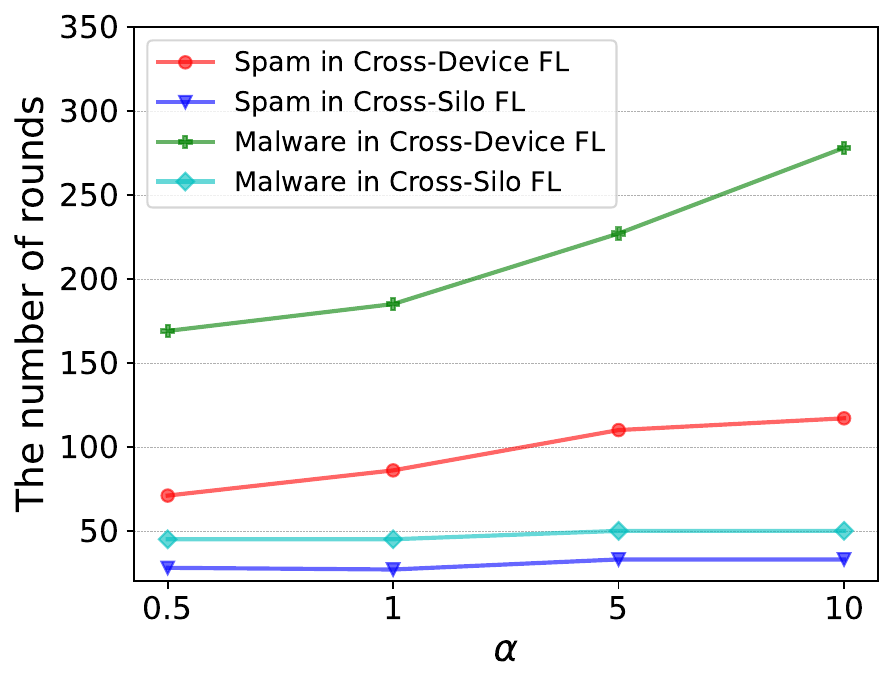}
    }
    \caption{The stats of FL-trained security classification models under various quantity-based non-IID data distributions. The $\alpha$ denotes the parameter of Dirichlet distribution.}
    \label{fig:fl_quantity_non-iid_stats}
\end{figure}

\finding{A high degree of quantity-based non-IID distribution can yield better model performance and faster training convergence, especially for cross-device FL.}

\subject{Quantity-based non-IID distribution.} The quantity-based non-IID is defined as the non-IID distribution of data samples across FL clients in terms of the quantity. It is assumed that these client-side local datasets still have the same label-wise distribution as the testing dataset on the FL server. The quantity of the client-side local dataset is abstracted as a Dirichlet distribution, and the extent of non-IID is configured through the parameter alpha $\alpha \in \{0.5, 1, 5, 10\}$ where the larger the $\alpha$, the less non-IID the respective data distribution is. For instance, in the case of $\alpha = 10$, FL clients are more likely to have same-sized local datasets compared with $\alpha = 1$. 

The performance statistics in quantity-based non-IID settings are listed in Table~\ref{tab:quantity_based_non_iid_spam} for SMS spam detection, and  Table~\ref{tab:quantity_based_non_iid_malware} for Android malware detection. Since the FL training process didn't converge stably in some quantity-based non-IID settings, we report here the average performance of the last 10 rounds of each FL training along with the standard deviation. 
As we can observe that the higher the quantity-based non-IID of client-side data is, the better performance the resulting FL model can achieve. In other words, as the extent of quantity-based non-IID increases with smaller $\alpha$, the resulting FL model achieves a better performance. For instance, when the extent of non-IID increases from $\alpha = 10$ to $\alpha = 0.5$,  the performance of the Android malware detection model has increased by 2.59\% (from 95.38\% to  97.97\%) for the  precision and by 0.23\% (from 99.53\% 99.76\%) for the recall. 
Such performance variance is also illustrated in Figure~\ref{fl_quantity_non-iid_accuracy}.

As presented in Figure~\ref{fl_quantity_non-iid_convergence}, another benefit of a higher extent of quantity-based non-IID is the fewer number of rounds it takes to converge in FL training.  Here, a patience of 10 is configured for spam and 3 for malware, when determining whether the training converges. 
Given such a benefit of quantity-based non-IID, one possible explanation is that a higher level of quantity-based non-IID leads to a greater variance in the local data size of the FL clients. Consequently, it allows some FL clients to train better local models more quickly. The aggregation of such early but better local model updates leads to faster convergence and better performance.
However, one exception is the SMS spam detection model trained through cross-silo FL, where the performance variance among different $\alpha$ does not exceed 0.001. Our explanation is that the average client-side SMS data size of each client selected in cross-silo FL is ten times larger than that in cross-device FL. As a result, regardless of the level of the quantity-based non-IID, the client-side SMS datasets in cross-silo FL are sufficiently large to generate a global model that can quickly converge with good performance. 
In a nutshell, it can be concluded that in practical scenarios, a higher level of quantity-based non-IID distribution may actually lead to better performance and faster convergence, especially for cross-device FL. However, on contrary, label-based non-IID can incur extra delay in convergence as well as making the training process unstable, as detailed below. 

\finding{A high degree of label-based non-IID distribution has no obvious impact on model performance, but can incur notable fluctuations (i.e., unstable convergence) in FL training, especially for the cross-device FL.}

\subject{Label-based non-IID distribution.} 
Under the label-based non-IID distribution, client-side local datasets vary widely in the label distribution. We still utilized the Dirichlet function to capture this distribution and experimented with different alpha values ($\alpha \in \{0.5, 1, 5, 10\}$). Specifically, given an $\alpha$, a Dirichlet sequence of $N$ values is generated to instruct how to distribute all the positive samples (spam samples or malware samples) to the $N$ clients, and another Dirichlet sequence of $N$ values with the same $\alpha$ is generated to specify what fraction of negative samples each FL client should account for. 

\begin{table}
    \centering
    \footnotesize
    \caption{The performance of FL-trained models under various \textbf{label-based} non-IID data distribution.}
    \label{tab:label_based_non_iid_malware}
    \begin{threeparttable}
        \begin{tabular}{cccc}
            \toprule
           FL & Dirichlet& Precision & Recall \\
           \midrule
           \multirow{4}{0.15\linewidth}{Spam Cross-Device} 
           & $\alpha = 0.5$ & $0.9811\pm0.0201$ &$0.9792\pm0.0116$\\
           &$\alpha = 1$ &$0.9773\pm0.0105$ &$0.9817\pm0.0063$\\
           &$\alpha = 5$ & $0.9831\pm0.0049$& $0.9633\pm0.0042$\\
           &$\alpha = 10$ & $0.9670\pm0.0067$ &$0.9680\pm0.0054$\\
           \midrule
            \multirow{4}{0.15\linewidth}{Spam Cross-Silo} 
           & $\alpha = 0.5$ & $0.9920\pm0.0011$ & $0.9865\pm0.0022$\\
           &$\alpha = 1$ & $0.9898\pm0.0016$& $0.9889\pm0.0021$\\
           &$\alpha = 5$ & $0.9898\pm0.0016$& $0.9866\pm0.0026$\\
           &$\alpha = 10$ & $0.9913\pm0.0011$ & $0.9896\pm0.0015$\\
             \midrule
           \multirow{4}{0.15\linewidth}{Malware Cross-Device} 
           & $\alpha = 0.5$ & $0.9611\pm0.0021$ & $1.000\pm0.0000$ \\
           &$\alpha = 1$ & $0.9689\pm0.0053$ & $0.9972\pm0.0009$ \\
           &$\alpha = 5$ & $0.9632\pm0.0027$ & $0.9974\pm0.0007$ \\
           &$\alpha = 10$ & $0.9541\pm0.0029$ & $0.9984\pm0.0015$ \\
           \midrule
            \multirow{4}{0.15\linewidth}{Malware Cross-Silo} 
           & $\alpha = 0.5$ & $0.9757\pm0.0011$ & $0.9821\pm0.0041$ \\
           &$\alpha = 1$ & $0.9521\pm0.0024$ & $0.9953\pm0.0000$ \\
           &$\alpha = 5$ & $0.9765\pm0.0056$ & $0.9953\pm0.0000$ \\
           &$\alpha = 10$ & $0.9607\pm0.0024$ & $0.9953\pm0.0000$ \\
           \bottomrule
        \end{tabular}
    \end{threeparttable}
\end{table}

\begin{figure}
    \centering
    \subfigure[SMS spam in cross-device FL.]{
        \label{fl_label_non_iid_convergence_spam_device}
        \includegraphics[width=.45\columnwidth]{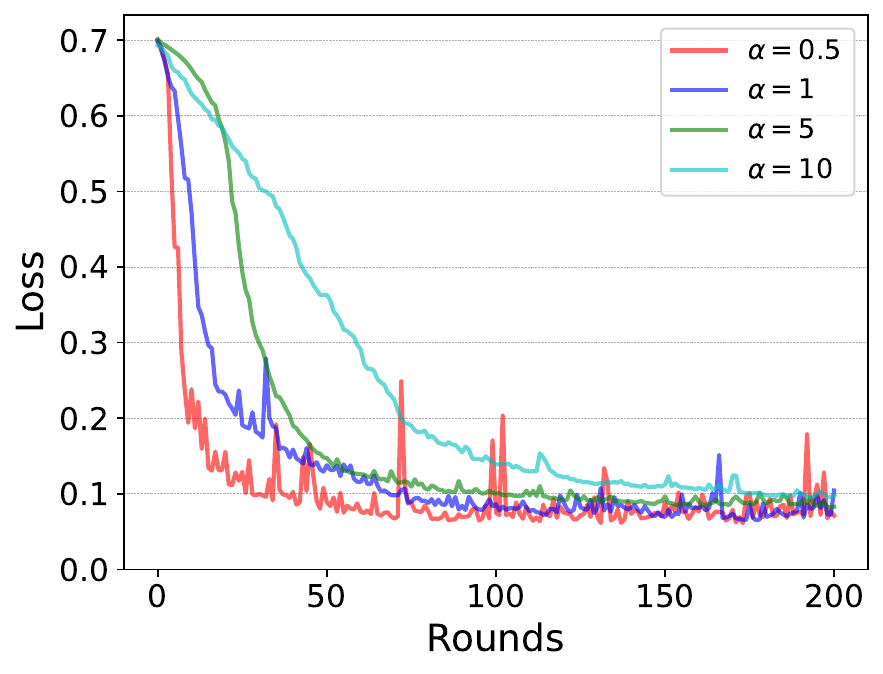}
    }
    \hfill
     \subfigure[SMS spam in cross-silo FL.]{
        \label{fl_label_non_iid_convergence_spam_silo}
        \includegraphics[width=.45\columnwidth]{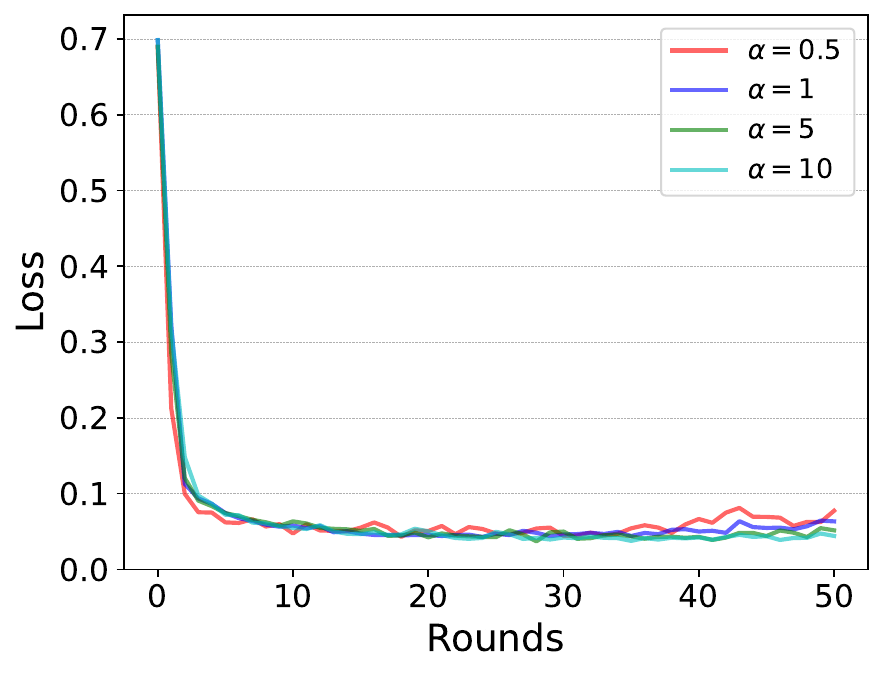}
    }
    \subfigure[Malware in cross-device FL.]{
        \label{fl_label_non_iid_convergence_malware_device}
        \includegraphics[width=.45\columnwidth]{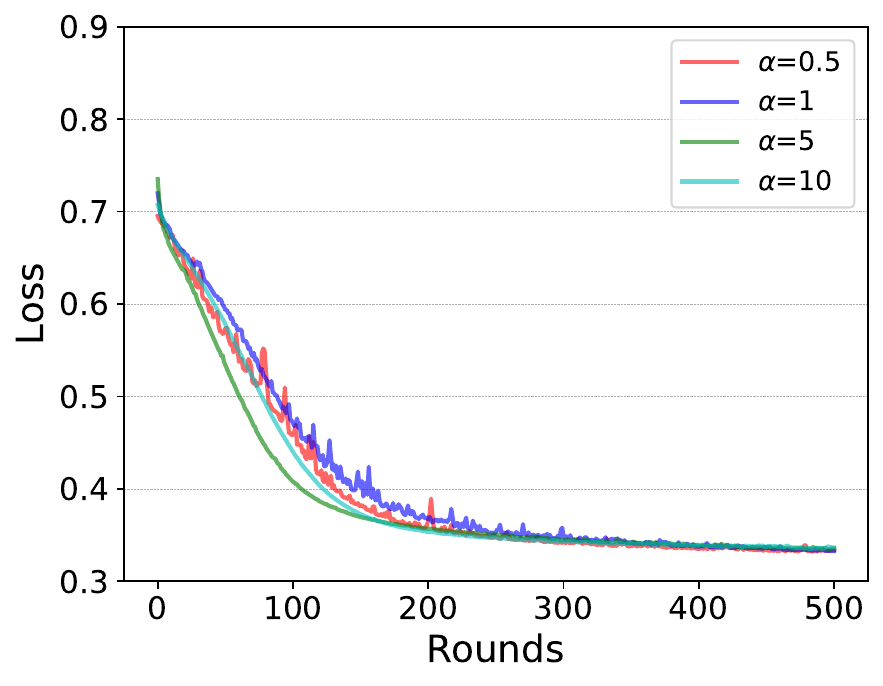}
    }
    \hfill
     \subfigure[Malware in cross-silo FL.]{
        \label{fl_label_non_iid_convergence_malware_silo}
        \includegraphics[width=.45\columnwidth]{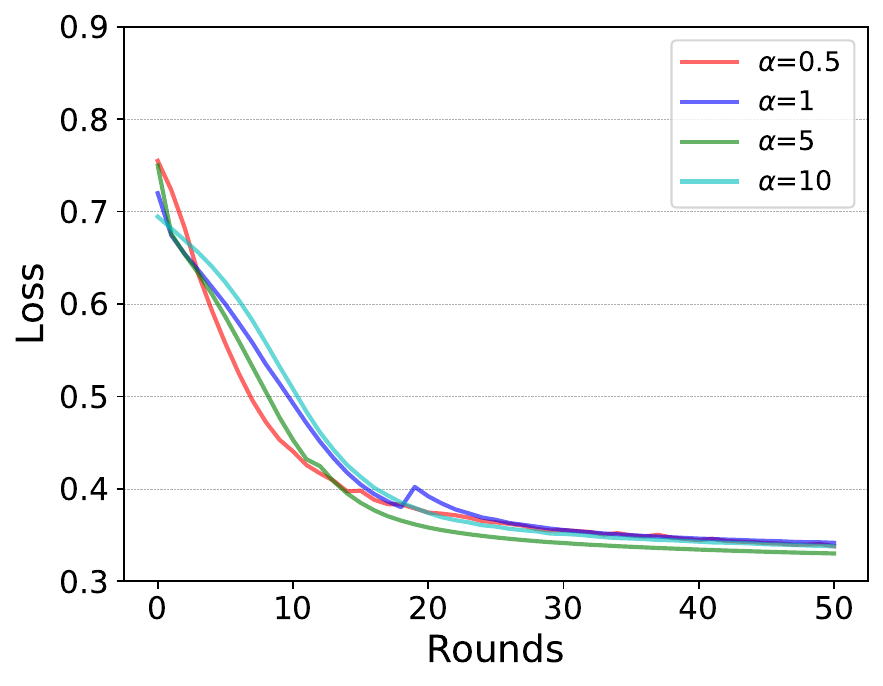}
    }
    \caption{The FL training processes for  label-based non-IID scenarios.}
    \label{fig:fl_label_non-iid_stats}
\end{figure}

Table~\ref{tab:label_based_non_iid_malware} lists the performance stats of the resulting FL-trained models under different $\alpha$ values, and we can see that the label-based non-IID distribution has no obvious impact on the model performance. Nevertheless, the higher the degree of label-based non-IID is, the more fluctuations we can observe during the FL training, which is likely due to client-side overfitting. For instance, in the cross-device FL, a SMS spam model trained with $\alpha = 0.5$ has a standard deviation of 2.01\% in its precision for the last 10 rounds of FL training, while it is only 0.67\% for $\alpha = 10$. 
As we can see, this phenomenon is especially obvious for cross-device FL, as further illustrated in Figure~\ref{fig:fl_label_non-iid_stats}.
We conclude that the label-based non-IID distribution can cause non-negligible instability (fluctuations) for the FL convergence process, especially for cross-device FL involving a large number of clients. Also, another side effect of label-based non-IID is the delay in convergence, which will be further elaborated in \S\ref{sec:efficiency}.

\finding{The consistent label imbalance (CLI) towards positive (malicious) samples can incur a non-negligible degradation in performance, while the performance impact of CLI towards negative samples (more practical in cross-device FL) is minor. }

\subject{Consistent label imbalance (CLI).} CLI refers to that FL clients \textit{consistently} have samples of one class outweighing samples of another, e.g., more benign apps than malware, or more non-spam messages than spam. Here, we define $PNR$, the ratio of positive samples to negative samples, to profile the extent of consistent label imbalance. Specifically, in cross-device FL, clients tend to have more negative samples than positive ones, i.e., $PNR < 1$. On the contrary, FL clients in the cross-silo scenario are more likely to have more positive samples than negative ones, which renders $PNR > 1$. For both scenarios, we have experimented $PNR \in \{0.25, 1, 4\}$. 

Given a $PNR$, the following strategies are designed to distribute samples to FL clients to guarantee that the clients' local datasets satisfy the given $PNR$. When $PNR \leq 1$, all the negative samples (except for the testing dataset) are first distributed to the $N$ FL clients by following a Dirichlet distribution with $\alpha = 1$, which assigns $NS_i$ negative samples to client $i$. Then, $\lceil NS_i \times PNR \rceil$ positive samples will be randomly selected from all the positive samples with replacement before being assigned to client $i$. Similarly, when $PNR > 1$, the positive samples for each client will first be assigned by following the same Dirichlet distribution. Then, given that the number of the positive samples for client $i$ is decided as $PN_i$,  the negative samples for client $i$ will be $NS_i = \lceil PN_i / PNR \rceil$, and they will be randomly sampled from the overall negative samples with replacement. 

\begin{table}
    \centering
    \footnotesize
    \caption{The accuracy of security models under CLI scenarios.}
    \label{tab:cli_accuracy}
    \begin{threeparttable}
        \begin{tabular}{ccccc}
            \toprule
           \multirow{2}{*}{Task} & \multirow{2}{*}{Model} & \multicolumn{3}{c}{$PNR$}\\
           && 0.25& 1 & 4\\
           \midrule
            \multirow{3}{*}{Spam} 
           & Central & $0.9865$ & $0.9898$& $0.9873$\\
           &Cross-Device FL &$0.9703$ &$0.9857$ &$0.9641$ \\
           &Cross-Silo FL & $0.9857$ & $0.9905$& $0.9821$\\
           \midrule
           \multirow{3}{*}{Malware} 
           & Central & $0.9854$ & $0.9963$ & $0.7866$ \\
           &Cross-Device FL & $0.9805$ & $0.9901$ & $0.5000$ \\
           &Cross-Silo FL & $0.9755$ & $0.9894$ & $0.5026$\\
           \bottomrule
        \end{tabular}
    \end{threeparttable}
\end{table}

\begin{figure}
    \centering
    \subfigure[SMS spam in cross-device FL.]{
        \label{fl_cli_convergence_spam_device}
        \includegraphics[width=.45\columnwidth]{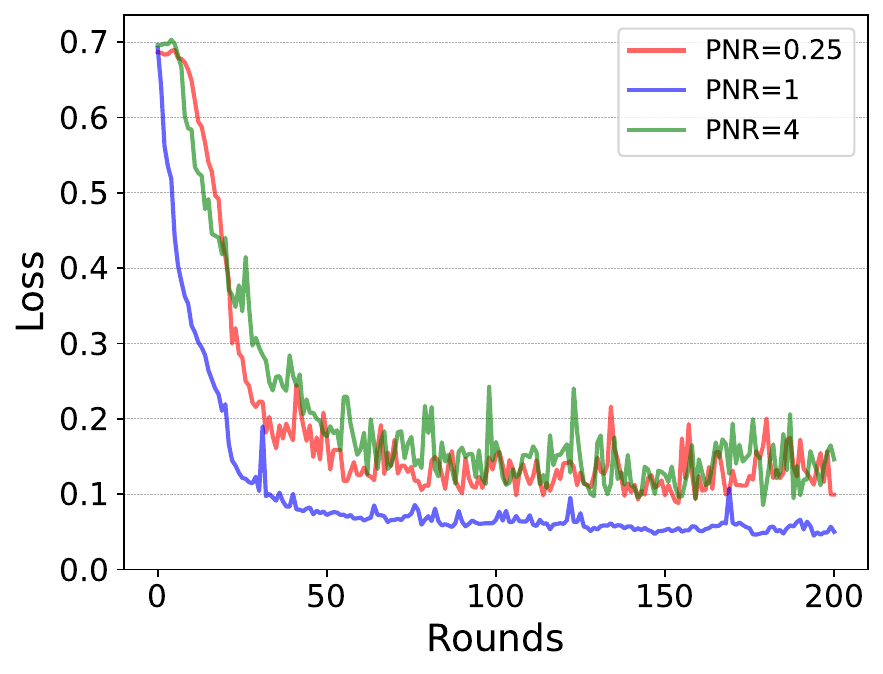}
    }
     \subfigure[SMS spam in cross-silo FL.]{
        \label{fl_cli_convergence_spam_silo}
        \includegraphics[width=.45\columnwidth]{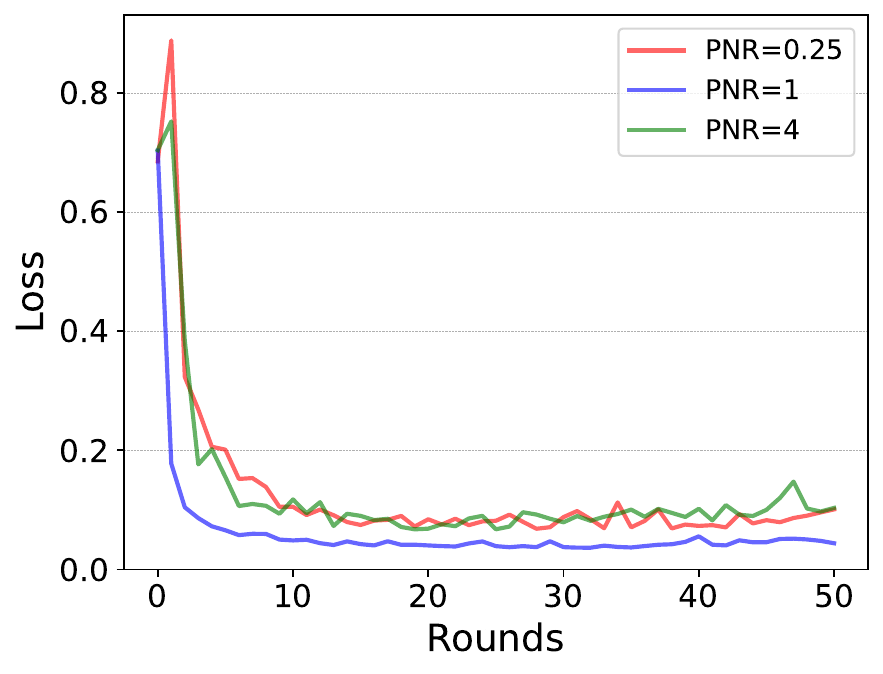}
    }
    \subfigure[Malware in cross-device FL.]{
        \label{fl_cli_convergence_malware_device}
        \includegraphics[width=.45\columnwidth]{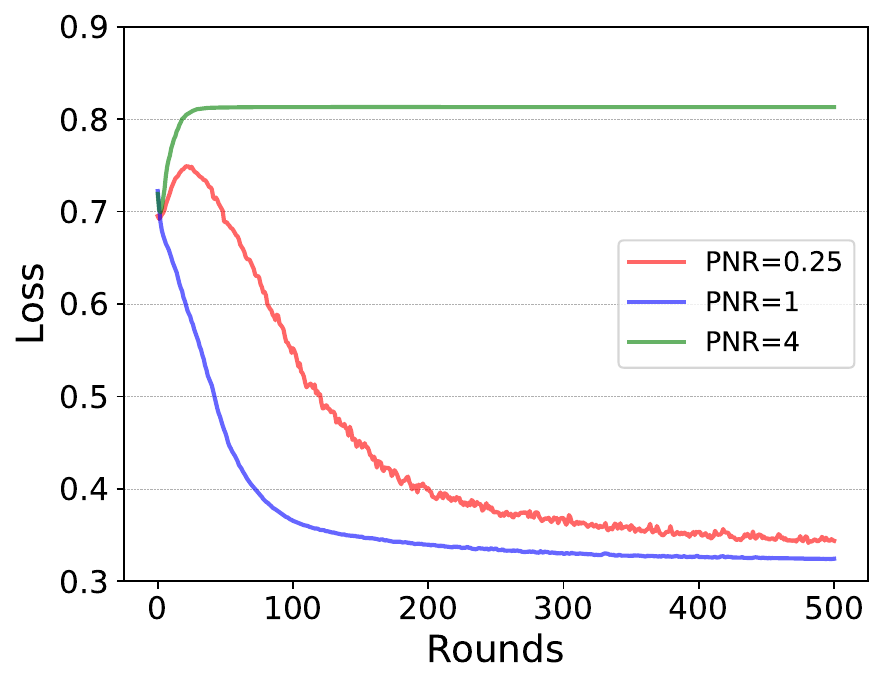}
    }
     \subfigure[Malware in cross-silo FL.]{
        \label{fl_cli_convergence_malware_silo}
        \includegraphics[width=.45\columnwidth]{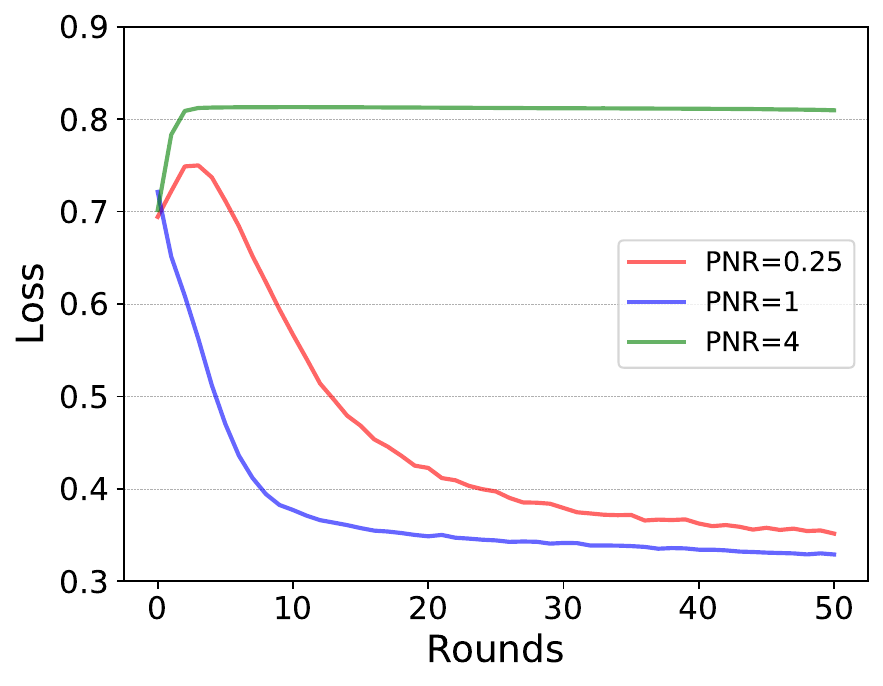}
    }
    \caption{The FL training processes for the scenarios of consistent label imbalance. }
    \label{fig:fl_cli_stats}
\end{figure}

Table~\ref{tab:cli_accuracy} presents the accuracy stats of the resulting models under different $PNR$ values. We can see that the balanced setting ($PNR = 1$) has achieved the best performance, across security tasks and FL scenarios. However, when the dataset is not balanced ($PNR \neq 1$), the cases with more positive samples ($PNR > 1$) can incur worse performance when compared to the cases with more negative samples ($PNR < 1$), especially for the cross-device scenario. Particularly, when the $PNR = 4$, the cross-device SMS spam model has achieved an accuracy of 96.41\%, which is 2.16\% lower than the balanced case and 0.62\% lower than $PNR = \frac{1}{4}$. 
This pattern is more obvious for the malware detection task which achieved an accuracy of 98.05\% at $PNR = \frac{1}{4}$ in cross-device FL while the counterpart at $PNR = 4$ didn't even converge (see Figure~\ref{fig:fl_cli_stats}). 
We thus conclude that the consistent label imbalance towards positive (malicious) samples can incur a non-negligible degradation in performance, while the impact of the CLI towards negative samples is minor. Further investigation shows that when $PNR > 1$, the FL-trained  model tends to generate much more false alarms, which leads to a much lower precision. For instance, when $PNR = 4$, compared with the centrally trained counterparts, FL-trained models have a similar recall but the precision has lowered by almost 4\% for the cross-device spam model and 29\% for the cross-device malware model. 
More details can be found in Appendix~\ref{appendix:cli_results}
As FL clients for cyber threat detection tend to have more negative samples than positive ones, we believe the CLI issue won't undermine the effectiveness of FL for cyber threat detection.

\subject{Language-based non-IID distribution for SMS spam detection.} When applying FL to SMS spam detection, FL clients, especially those in the cross-device scenario, are more likely to have all their spam/non-spam messages composed in very few natural languages. It is thus interesting and necessary to evaluate the impact of such a language-based non-IID distribution on FL effectiveness. As detailed in \S\ref{sec:security_tasks}, samples in our spam/non-spam dataset are of 25 different natural languages. In our language-relevant experiments, each FL client will first be randomly assigned with $k$ different languages, which means this client can only have samples belonging to one of these $k$ languages.  Then, samples of language $l$ will be assigned only to the subset of clients that qualify this language requirement, and a Dirichlet distribution with $\alpha = 1$ is used to fulfill the sample assignment. In our experiments, 4 different $k$ values have been evaluated, which are $k \in \{1, 2, 3, 4\}$.  The respective performance stats are listed in Table~\ref{tab:lang_based_non_iid_spam}.
In a nutshell, the more languages each FL client is assigned with, the better precision the respective FL-trained model can achieve, while the effect of language-based non-IID on the recall is minor. 

\begin{table}
    \centering
    \footnotesize
    \caption{The performance stats of FL-trained spam detection models under language-based non-IID distribution.}
    \label{tab:lang_based_non_iid_spam}
    \begin{threeparttable}
        \begin{tabular}{cccc}
            \toprule
           FL & $k$ & Precision & Recall \\
           \midrule
           \multirow{4}{0.15\linewidth}{Cross-Device} 
           & $k = 1$ & $0.9609\pm0.0087$ & $0.9889\pm0.0025$ \\
           &$k = 2$ & $0.9726\pm0.0059$ & $0.9896\pm0.0022$\\
           &$k = 3$ & $0.9771\pm0.0101$& $0.9861\pm0.0042$\\
           &$k = 4$ & $0.9796\pm0.0063$ & $0.9891\pm0.0024$\\
           \midrule
           \multirow{4}{0.15\linewidth}{Cross-Silo} 
           & $k = 1$ & $0.9760\pm0.0037$ & $0.9162\pm0.0071$\\
           &$k = 2$ &$0.9850\pm0.0044$ & $0.9920\pm0.0019$\\
           &$k = 3$ & $0.9840\pm0.0042$& $0.9901\pm0.0017$\\
           &$k = 4$ &$0.9866\pm0.0044$  & $0.9895\pm0.0025$\\
           \bottomrule
        \end{tabular}
    \end{threeparttable}
\end{table}

\begin{figure}
    \centering
    \subfigure[Malware in cross-device FL.]{
        \label{fl_malware_family_convergence_malware_device}
        \includegraphics[width=.45\columnwidth]{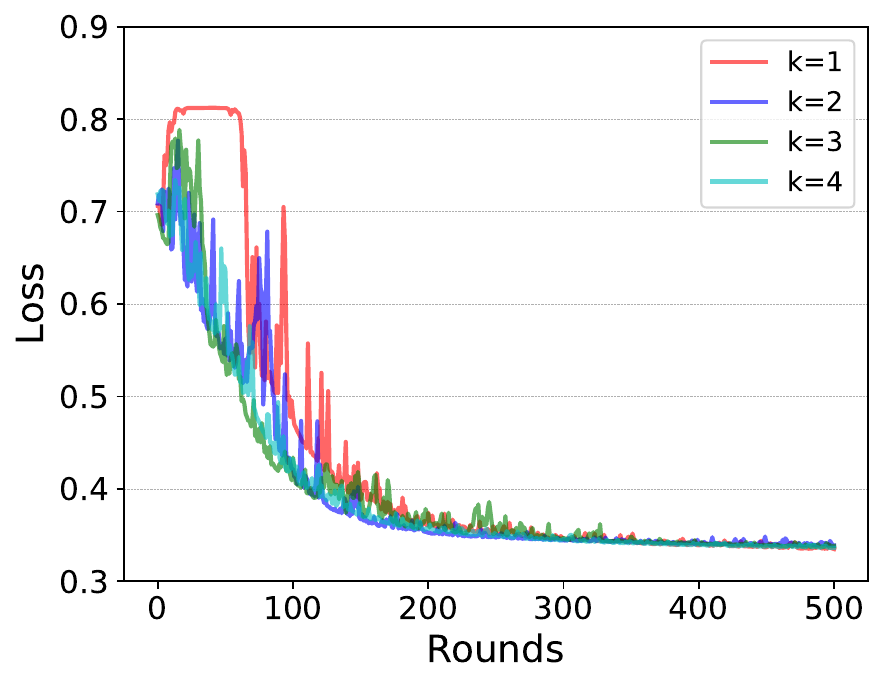}
    }
     \subfigure[Malware in cross-silo FL.]{
        \label{fl_malware_family_convergence_malware_silo}
        \includegraphics[width=.45\columnwidth]{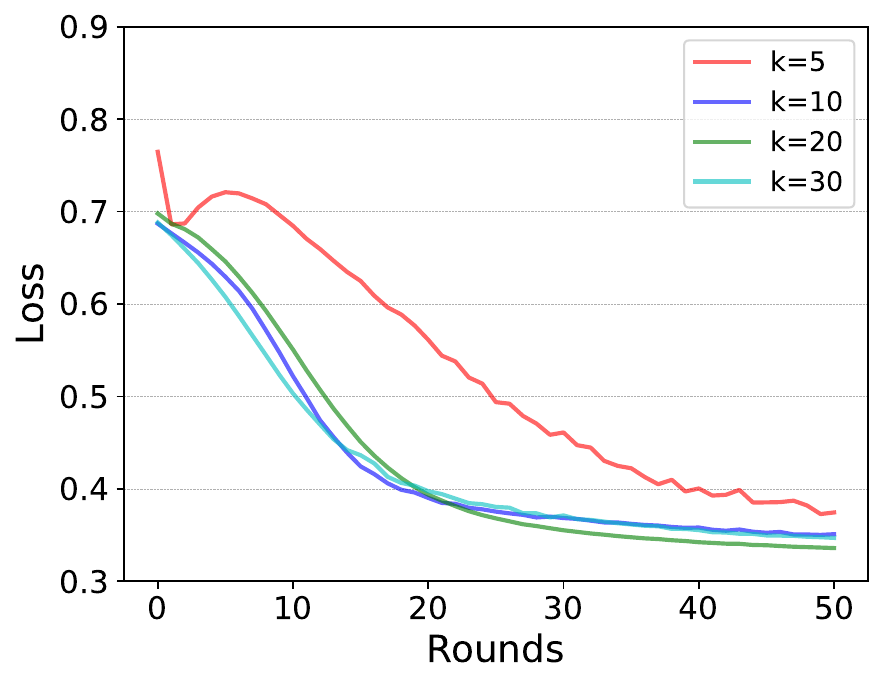}
    }
    \caption{The FL training processes for the scenarios of family imbalance for malware detection.}
    \label{fig:fl_lang_family_non-iid_stats}
\end{figure}

\subject{Family-based non-IID distribution for Android malware detection.} Similar to spam detection, malware detection also has a unique non-IID distribution that is related to malware families. Specifically, there exists an imbalanced distribution of malware families across FL clients, especially in the scenario of cross-device FL. To profile its potential impact on FL training, we conducted experiments following a setting that is similar to that of aforementioned language-based non-IID experiments. Here, we have 133 distinct malware families. Considering cross-silo FL only has 20 clients in total, in order not to miss out too much families, each cross-silo client needs to select at least $\{k:k>4\}$ families. The following $k$ values have been explored: $\{ 5, 10, 20, 30\}$ in cross-silo FL and $\{ 1, 2, 3, 4\}$ in cross-device FL. 
    
The respective performance stats can be found in Table~\ref{tab:family_based_non_iid_malware} while the training process is illustrated in Figure~\ref{fig:fl_lang_family_non-iid_stats}. We can see that, as $k$ gets smaller, the training processing becomes more unstable and converges more slowly. However there is no obvious pattern observed for the model performance, which is different from the language-based non-IID distribution for SMS spam detection.

\begin{table}
    \centering
    \footnotesize
    \caption{The performance stats of FL-trained \textbf{malware detection models} under family-based non-IID distribution.}
    \label{tab:family_based_non_iid_malware}
    \begin{threeparttable}
        \begin{tabular}{cccc}
            \toprule
           FL & $k$ & Precision & Recall \\
           \midrule
           \multirow{4}{0.15\linewidth}{Cross-Device} 
           & $k = 1$ & $0.9617\pm0.0080$ & $0.9972\pm0.0027$ \\
           &$k = 2$ & $0.9503\pm0.0053$ & $0.9967\pm0.0056$\\
           &$k = 3$ & $0.9530\pm0.0050$ & $0.9967\pm0.0012$ \\
           &$k = 4$ & $0.9615\pm0.0044$ & $0.9976\pm0.0000$ \\
           \midrule
           \multirow{4}{0.15\linewidth}{Cross-Silo} 
           & $k = 5$ & $0.9774\pm0.0018$ & $0.9548\pm0.0090$ \\
           &$k = 10$ & $0.9472\pm0.0025$ & $0.9788\pm0.0033$ \\
           &$k = 20$ & $0.9742\pm0.0009$ & $0.9951\pm0.0007$ \\
           &$k = 30$ & $0.9609\pm0.0016$ & $0.9875\pm0.0018$ \\
           \bottomrule
        \end{tabular}
    \end{threeparttable}
\end{table} 

\finding{Non-IID data distribution can lead to high instability during the convergence of the respective FL training process.}

\begin{figure}
    \centering
    \subfigure[Spam in cross-device FL.]{
        \label{fl_unstable_spam_device}
        \includegraphics[width=.45\columnwidth]{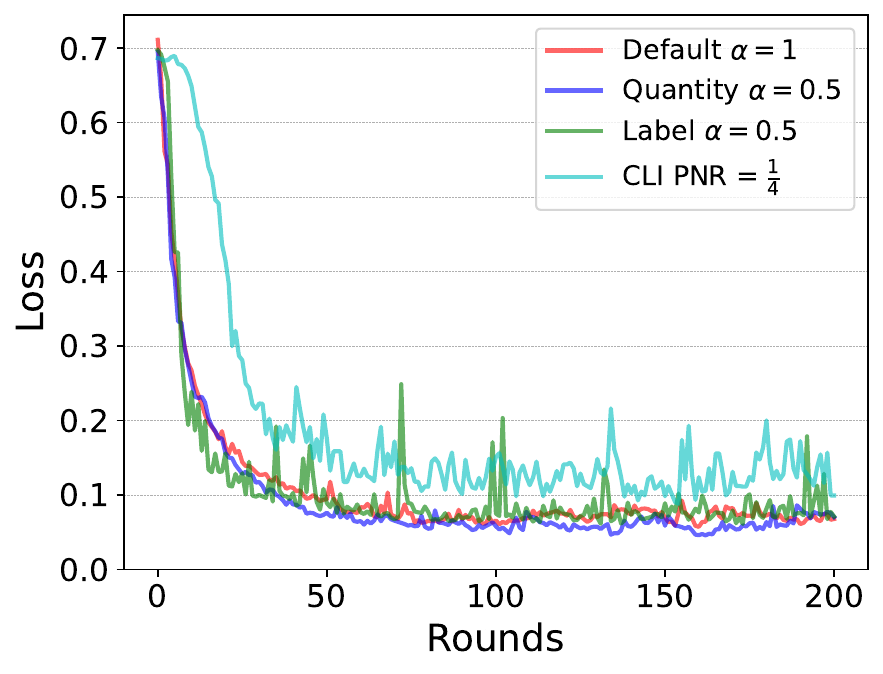}
    }
     \subfigure[Spam in cross-silo FL.]{
        \label{fl_unstable_spam_silo}
        \includegraphics[width=.45\columnwidth]{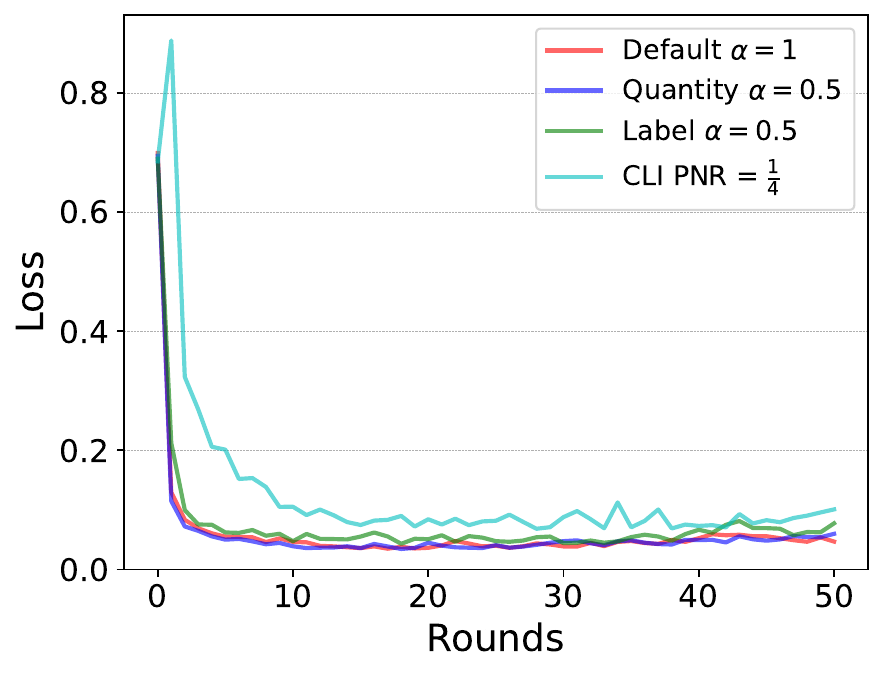}
    }
    \subfigure[Malware in cross-device FL.]{
        \label{fl_unstable_malware_device}
        \includegraphics[width=.45\columnwidth]{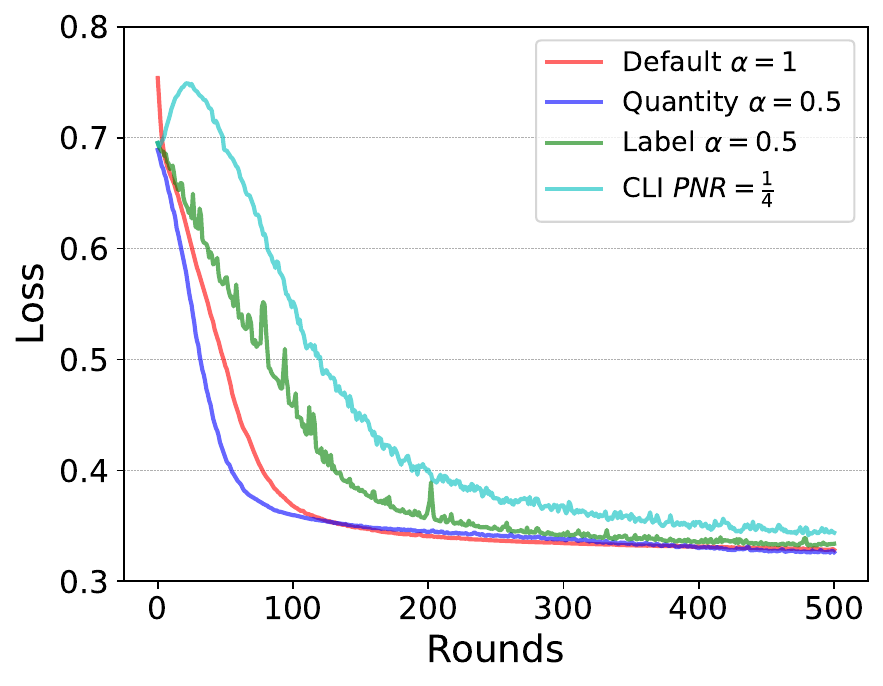}
    }
     \subfigure[Malware in cross-silo FL.]{
        \label{fl_unstable_malware_silo}
        \includegraphics[width=.45\columnwidth]{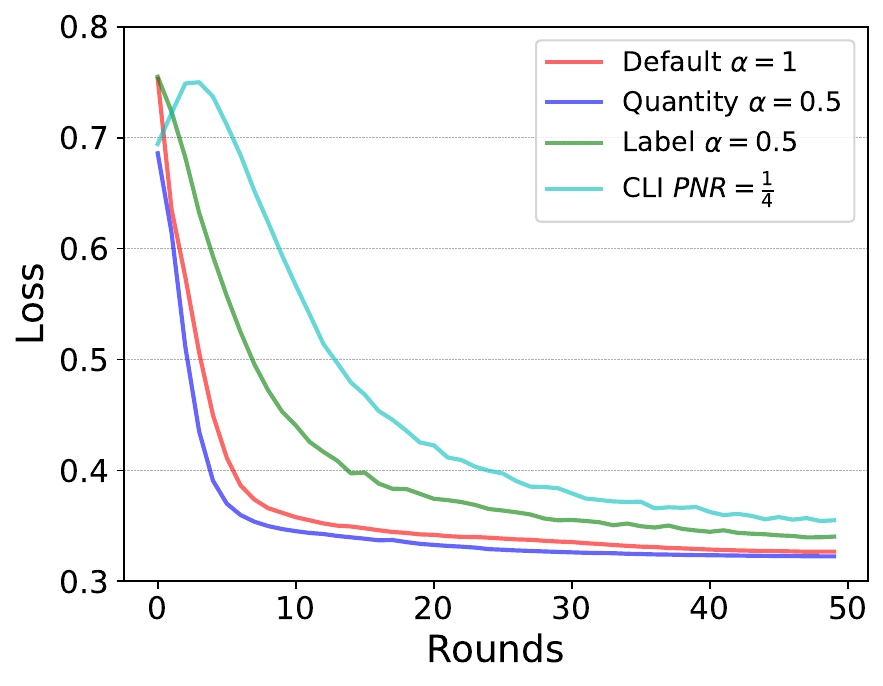}
    }
    \caption{The (in)stability of the FL training processes for the scenarios of extreme non-IID distribution.  }
    \label{fig:fl_unstable_convergence}
\end{figure}

\subject{The convergence of FL training in non-IID or imbalanced scenarios.} As revealed from above experiments, some non-IID scenarios lead to higher instability for the convergence of the respective FL training process. For instance, in the scenario of label-based non-IID distribution with $\alpha=0.5$, given the models trained during the last 10 rounds, the standard deviation of their respective accuracy is 0.81\% while it is only 0.07\% for our baseline model for SMS spam. This observation is further illustrated in Figure~\ref{fig:fl_unstable_convergence} showing the FL training process for various non-IID distributions. We can see the consistent label imbalance with $PNR = \frac{1}{4}$ stands out to suffer from the most severe convergence instability, which is followed by the label-based non-IID with $\alpha = 0.5$ and quantity-based non-IID with $\alpha = 0.5$.

We have also defined a set of metrics so as to quantify the instability of the convergence process across aforementioned non-IID scenarios, which distills the same observation as Figure~\ref{fig:fl_unstable_convergence}. For more details, please refer to Appendix~\ref{appendix:convergence_instability}. To address the instability issue, we proposed and evaluated the bootstrapping strategy which will be detailed in Section~\ref{sub:bootstrapping}.

%% file: 5_1_adversarial_robustness.tex
In this section, we move to evaluate the byzantine resilience of FL with regards to data poisoning and model poisoning. 

\subject{The threat model.} When selecting data/model poisoning attacks and evaluating their attack impact, we consider a realistic threat model. Specifically, for both data poisoning and model poisoning, we assume that
the attacker has whitebox knowledge of the global model (its parameters, architecture, and output), as this knowledge can be easily inferred from either the model updates or the local training process. However, the attacker has no knowledge of the aggregation rule that is enforced by the central server, so any attacks that rely on such knowledge are considered as impractical and are outside the scope of our evaluation. 

Also, for both cross-device FL and cross-silo FL, the attackers should have no access to benign FL clients, regardless of data poisoning or model poisoning. 
Besides, we assume that data poisoning attacks don't involve any effective collusion among poisoned clients as the attackers of data poisoning either have no direct access to the poisoned FL clients or the access is too limited to allow effective collusion. Instead, a model poisoning attacker in the cross-device FL should have compromised one or more FL clients and these compromised FL clients can collude with each other during the FL training process. 
Note that such a model poisoning attack is  considered only possible for the cross-device FL but not the cross-silo FL, as the participants of the cross-silo FL are assumed to be trusted organizations under contracts and their FL servers should be well protected. Regarding to the ratio of poisoned FL clients $M$, we consider $M \leq 5\%$ as practical and realistic for data poisoning while it is $M \leq \%1$  for model poisoning, which is aligned with recent works on practical poisoning attacks~\cite{shejwalkar2022back}. Then, in terms of the attack goals, the attackers aim to degrade the overall performance of the poisoned model and cause the model to misclassify any given test input. 

To summarize, we consider untargeted FL poisoning attacks and the attackers have whitebox knowledge of the global model, but no knowledge of the server-side aggregation rule. Then, a practical data poisoning attack is equipped with very limited access to the poisoned FL clients, and thus does not involve cross-client collusion during FL training, whereas a practical model poisoning attack has full access of the poisoned FL clients to enable cross-client collusion, but is only feasible for the cross-device FL.

\subsection{Data Poisoning Attacks}
\finding{Data poisoning with a practical fraction of poisoned clients ($M \leq 5\%$) has a \textit{negligible} attack impact in terms of accuracy decrease, and even poisoning with an impractically high $M = 40\%$ still results in minor attack impact as long as the per-client sample poisoning rate is realistic (e.g., 25\%). }

\subject{Data poisoning under the default FL setting.} We first evaluated the untargeted data poisoning attacks under the default FL setting (Table~\ref{tab:default_fl_setting}) where the AGR is the FedAVG and the cross-device data distribution is quantity-based Dirichlet with $\alpha = 1$. Here, we define $M$ as the fraction of \textit{manipulated FL clients} that have their local datasets been poisoned to some extent. Also, for each manipulated FL client,  the fraction of local samples being poisoned is defined as the poisoning ratio $p$.
For instance, when $p = 0.5$, each of the manipulated FL clients ($N \times M$) has 50\% of its local datasets been manipulated. Then, regarding the manipulation strategy, we consider the static label flipping, which means the manipulated samples have its payload unchanged but its label flipped. For instance, a spam message under manipulation has its label switched to non-spam but its text content unchanged. Although the settings of different $M$ has been explored in multiple studies~\cite{tolpegin2020data, shejwalkar2022back} on data poisoning, \textit{we are the first to explore the impact of different poisoning rate $p$}, which is essential to our research goal of evaluating FL in \textit{realistic} scenarios. 

\begin{figure}
    \centering
    \subfigure[Spam.]{
        \label{fig:fl_data_poisoning_rate_default_M_setting_spam_cross_device}
        \includegraphics[width=.45\columnwidth]{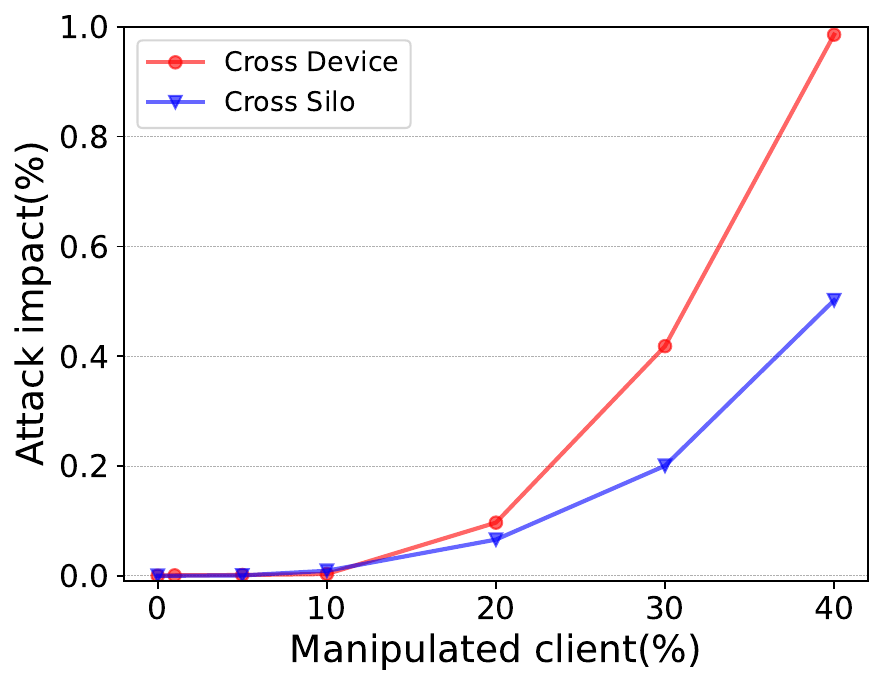}
    }
    \subfigure[Malware.]{
        \label{fig:fl_data_poisoning_rate_default_M_setting_malware_cross_device}
        \includegraphics[width=.45\columnwidth]{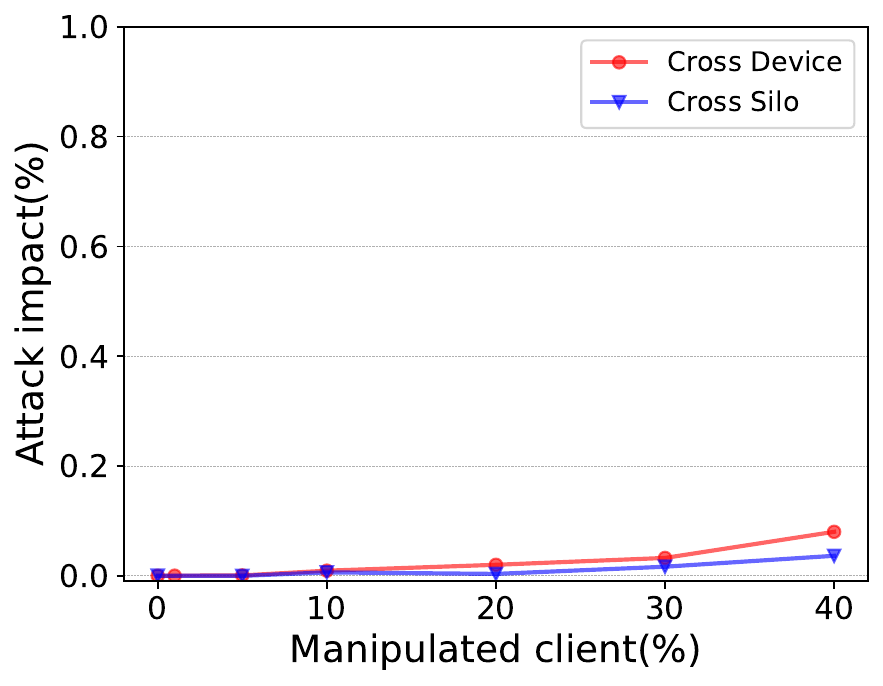}
    }

    \caption{The attack impact in terms of accuracy decrease for data poisoning attacks of different fractions of manipulated clients $M$, when $p=100\%$.} 
    \label{fig:fl_data_poisoning_rate_default_M_setting}
\end{figure}

We first explored the attack impact of different fractions of manipulated clients $M \in \{1\%, 5\%, 10\%, 20\%, 30\%, 40\%\}$ at a poisoning rate of $p = 100\%$. 
The attack results are presented in Table~\ref{tab:default_SLF_data_poison} and Figure~\ref{fig:fl_data_poisoning_rate_default_M_setting} while the corresponding FL training processes are illustrated in Appendix~\ref{appendix:data_poisoning_training}. 
As we can see, when $M$ is in the practical range ($M \leq 5$\%) of data poisoning, the impact of the attack on accuracy degradation is less than 0.2\%, and thus negligible.  When $M$ is set to a large ratio, e.g., $20\%$, the attack impact can be significant, 9.68\% for spam detection and 2.00\% for malware detection, which is consistent with previous results on non-security FL tasks~\cite{tolpegin2020data, shejwalkar2022back}. Also, compared to spam detection, malware detection models appear to be more robust in terms of defending against large but impractical fractions of manipulated clients. For instance, when $M = 30\%$, the data poisoning attack has achieved an accuracy degradation of 41.83\% on SMS spam detection while it is only 3.25\% for Android malware detection. 
\begin{table}
    \centering
    \footnotesize
    \caption{The attack impact in terms of accuracy decrease for \textbf{data poisoning attacks} at different fractions of manipulated clients $M$ and the same local data poisoning rate $p = 100\%$.}
    \label{tab:default_SLF_data_poison}
    \begin{threeparttable}
        \begin{tabular}{ccccc}
            \toprule
            Task & $M$ & Cross-device & Cross-silo\\
            \midrule
            \multirow{6}{*}{Spam}
                & 1\% & $0.0008\pm0.0009$& -\tnote{1}\\
                & 5\% & $0.0014\pm0.0020$& $0.0006\pm0.0009$\\
                & 10\%& $0.0032\pm0.0024$& $0.0091\pm0.0178$\\
                & 20\%& $0.0968\pm0.1453$& $0.0659\pm0.1214$\\
                & 30\%& $0.4183\pm0.1164$& $0.2004\pm0.1573$\\
                & 40\%& $0.9858\pm0.0007$ & $0.5020\pm0.0153$\\
             \midrule
            \multirow{6}{*}{Malware}
                & 1\% & $0.0002\pm0.0005$ & -\tnote{1} \\
                & 5\% & $0.0006\pm0.0008$ & $-0.0001\pm0.0013$\\
                & 10\% & $0.0093\pm0.0008$ & $0.0062\pm0.0018$\\
                & 20\% & $0.0200\pm0.0064$ & $0.0034\pm0.0010$\\
                & 30\% & $0.0325\pm0.0341$ & $0.0165\pm0.0009$\\
                & 40\% & $0.0801\pm0.0398$ & $0.0364\pm0.0041$\\
           \bottomrule
        \end{tabular}
        \begin{tablenotes}
            \item [1] In cross-silo FL, there's no $M=1\%$ result, since 1 compromised client corresponds to $M=5\%$.
        \end{tablenotes}
    \end{threeparttable}
\end{table}

\begin{table}
    \centering
    \footnotesize
    \caption{The attack impact in terms of accuracy decrease for \textbf{data poisoning attacks} with $M = 40\%$ and different poisoning rates $p$.}
    \label{tab:default_DLF_data_poison_different_p}
    \begin{threeparttable}
        \begin{tabular}{ccc}
            \toprule
            Task & $p$ & Cross-device\\
            \midrule
            \multirow{4}{*}{Spam}
                & 25\% & $0.0081\pm0.0038$\\
                & 50\% & $0.0402\pm0.0325$\\
                & 75\% & $0.4579\pm0.0505$\\
                & 100\%& $0.9858\pm0.0007$\\
             \midrule
            \multirow{4}{*}{Malware}
                & 25\% & $0.0121\pm0.0032$\\
                & 50\% & $0.0126\pm0.0107$\\
                & 75\% & $0.0154\pm0.0039$\\
                & 100\%& $0.0927\pm0.0293$\\
           \bottomrule
        \end{tabular}
    \end{threeparttable}
\end{table}

As the effect of data poisoning is negligible when $M \leq 5$ and $p = 100\%$, we didn't proceed to evaluate lower $p$ for $M \leq 5$.  Instead, we explored the effect of different $p$ at $M = 40\%$, for which $p \in \{ 25\%, 50\%, 75\%, 100\% \}$ were considered. The attack results are listed in Table~\ref{tab:default_DLF_data_poison_different_p}. As we can see, even if the $M$ is impractically high at $40\%$, a practical sample poisoning rate of $p = 25\%$ renders the data poisoning almost ineffective, which decreases the model accuracy by only 0.81\% for SMS spam detection and 1.21\% for Android malware detection. We also considered the scenario wherein security vendors with large datasets may all have their data poisoned but at a low fraction. Specifically, we further explored the cross-silo FL at $M = 100\%$ and $p = 10\%$, for which, the attack impact turns out to be negligible.

\subject{Data poisoning under label-based non-IID data distribution.} As aforementioned, under the default FL setting, practical data poisoning attacks have a negligible attack impact in terms of model performance degradation. We thus moved on to explore whether they can work in the label-based non-IID scenario, which is a more realistic FL scenario and may thus give the attackers more poisoning opportunities.  
Specifically, we consider two practical attack settings, one with $M = 5\%, p = 50\%$ and the other with $M = 5\%, p = 100\%$. We then evaluated their attack impact under different label-based non-IID distributions which vary by the Dirichlet parameter $\alpha = \{0.5, 1, 5, 10\}$. And we observed that data poisoning has no attack impact across these label-based non-IID scenarios and practical poisoning settings, which is consistent with our observation for the default FL settings. 
Specific data points can be found in Appendix~\ref{appendix:data_poison_label_non_idd}.

\begin{table}
    \centering
    \footnotesize
    \caption{A comparison among different AGRs in terms of defending against \textbf{data poisoning attacks} with impractically high $M$.  }
    \label{tab:default_SLF_data_poison_AGR}
    \begin{threeparttable}
        \begin{tabular}{ccccc}
            \toprule
            Task & $M$ & Trimmed Mean & Multi-Krum & FedAVG\\
            \midrule
            \multirow{3}{*}{Spam}
                & 20\%& $0.0269$~\tnote{1}& $0.0129$& $0.0968$\\
                & 30\%& $0.0510$& $0.0239$& $0.4183$\\
                & 40\%& $0.4458$& $0.2231$& $0.9858$\\
             \midrule
            \multirow{3}{*}{Malware}
                & 20\%& $0.0169$ & $0.0174$ & $0.0200$\\
                & 30\%& $0.0091$ & $0.0173$ & $0.0325$\\
                & 40\%& $0.0261$ & $0.0216$ & $0.0801$\\
           \bottomrule
        \end{tabular}
        \begin{tablenotes}
            \item [1] Each cell denotes the attack impact in terms of the loss of model accuracy.
        \end{tablenotes}
    \end{threeparttable}
\end{table}

\begin{figure}
    \centering
    \footnotesize
    \subfigure[Spam with Trimmed mean.]{
        \label{fig:fl_data_poisoning_trimmed_spam}
        \includegraphics[width=.45\columnwidth]{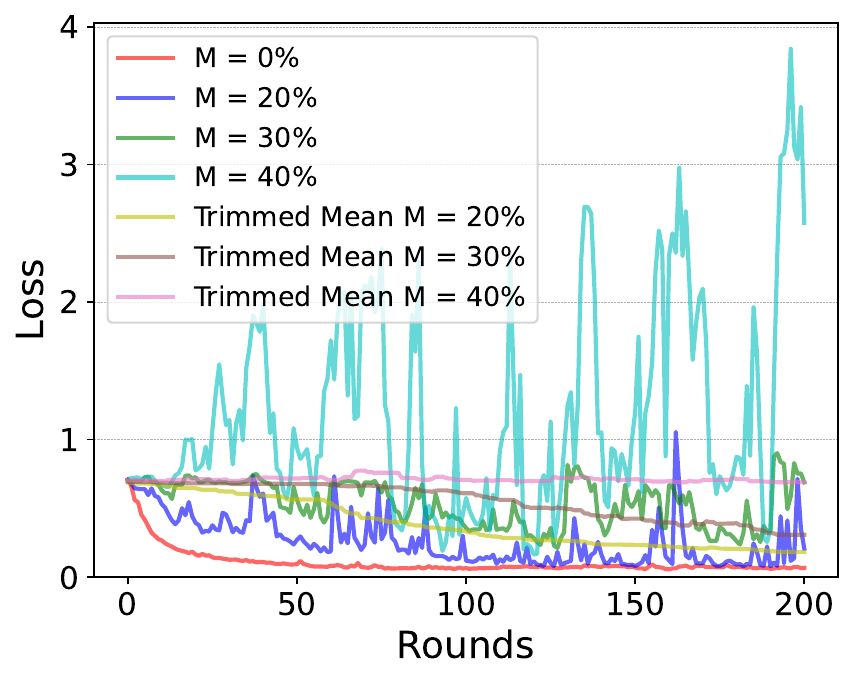}
    }
    \subfigure[Spam with Multi-Krum.]{
        \label{fig:fl_data_poisoning_multikrum_spam}
        \includegraphics[width=.45\columnwidth]{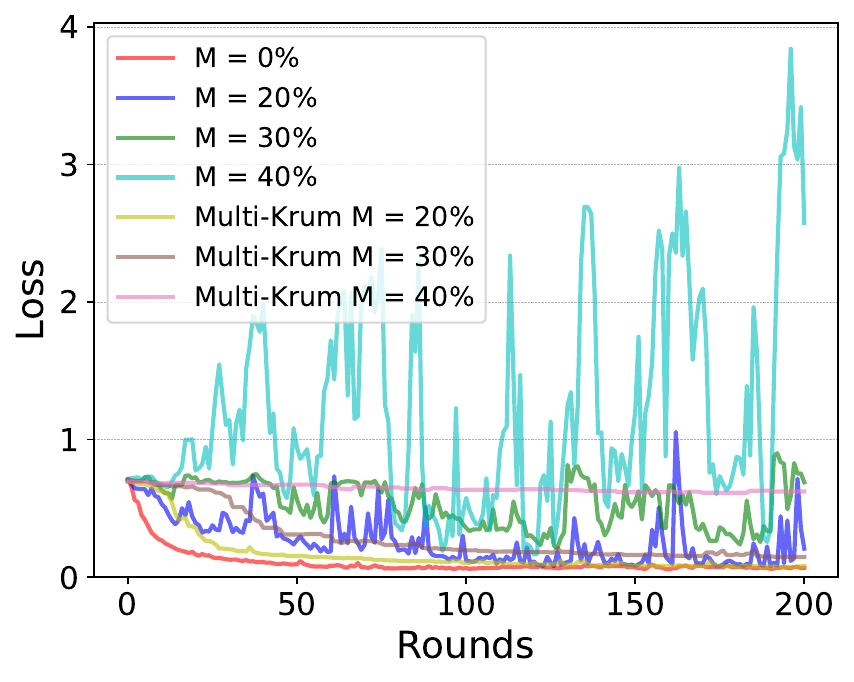}
    }
    \subfigure[Malware with Trimmed mean.]{
        \label{fig:fl_data_poisoning_trimmed_malware}
        \includegraphics[width=.45\columnwidth]{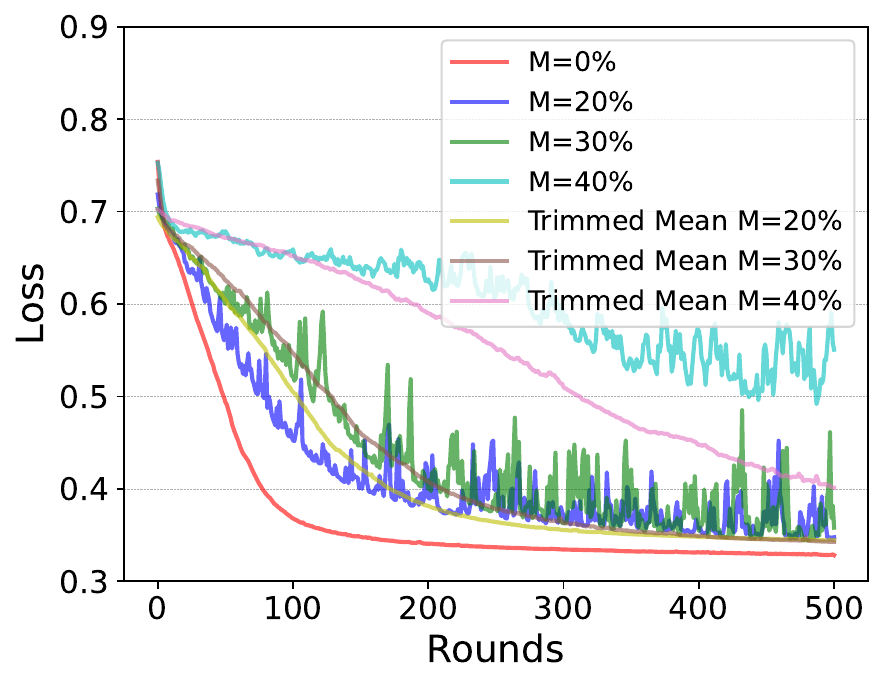}
    }
    \subfigure[Malware with Multi-Krum.]{
        \label{fig:fl_data_poisoning_multikrum_malware}
        \includegraphics[width=.45\columnwidth]{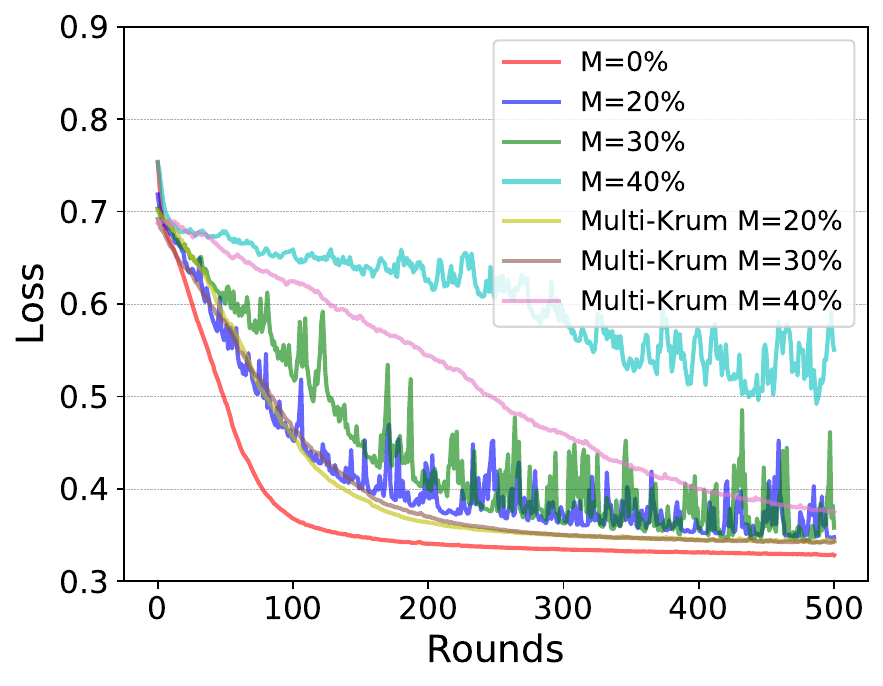}
    }

    \caption{The training processes under data poisoning attacks in cross-device FL with or without robust AGRs enforced.}
    \label{fig:fl_data_poisoning_AGR}
\end{figure}

\finding{Given impractically powerful data poisoning attackers (equipped with impractically high $M$ and $p$), robust AGRs can not only significantly lower the attack impact but also make the training process more smooth.}

\subject{Data poisoning under robust AGRs.} Despite impractical, some data poisoning attacks with both high $M$ and high $p$ still achieve non-negligible attack impact, so we moved on to evaluate the effectiveness of robust AGRs in terms of defending against such types of impractical data poisoning attacks. Specifically, two representative robust AGRs (Trimmed Mean~\cite{pmlr-v80-yin18a} and Multi-Krum~\cite{NIPS2017}) have been tested in our experiments.
Table~\ref{tab:default_SLF_data_poison_AGR} presents a direct comparison among these AGRs in terms of preventing data poisoning attacks, and we can observe that robust AGRs can effectively prevent data poisoning attacks to some extent. For instance, when $M = 30\%$, the attack on FedAVG resulted in a 41.83\% decrease in accuracy for SMS spam detection, but this decreased to 2.39\% with the use of Multi-Krum and to 5.10\% with Trimmed Mean.  
Figure~\ref{fig:fl_data_poisoning_AGR} also presents a comparison between training processes with robust AGRs and training processes with the default FedAVG. We can see that robust AGRs can not only help mitigate impractical but effective data poisoning attacks, but also smooth the training process, i.e., making the training process converge more stably.

\subsection{Model Poisoning Attacks}

\finding{Model poisoning with a practical fraction of poisoned clients ($M \leq 1\%$) has an either negligible or minor attack impact for cyber threat detection tasks.}

\begin{table}
    \centering
    \footnotesize
    \caption{The attack impact in terms of accuracy decrease for \textbf{model poisoning attacks} in the default FL settings.}
    \label{tab:pratical_model_poison_default_fl}
    \begin{threeparttable}
        \begin{tabular}{ccccc}
            \toprule
            Task & $M$ & LIE & MIN-MAX & MIN-SUM \\
            \midrule
            \multirow{6}{*}{Spam}
                & 1\% & $0.0003$& $ 0.0152$& $0.0130$\\
                & 5\% & $0.0016$& $0.0177$& $0.0166$\\
                & 10\%& $0.0014$& $0.0175$& $0.0164$\\
                & 20\%& $0.0019$& $0.0180$& $0.0168$\\
                & 30\%& $0.0014$& $0.0197$& $0.0241$\\
                & 40\% & $0.0026$ & $0.0201$& $0.0199$\\
             \midrule
            \multirow{6}{*}{Malware}
                & 1\% & $0.0002$ & $0.0014$ & $-0.0022$\\
                & 5\% & $0.0054$ & $-0.0007$ & $0.0073$\\
                & 10\% & $-0.0038$ & $0.0049$ & $-0.0039$\\
                & 20\% & $-0.0007$ & $0.0042$ & $0.0049$\\
                & 30\% & $0.0048$ & $0.0039$ & $0.0053$\\
                & 40\% & $0.0065$ & $0.0015$ & $0.0176$\\
           \bottomrule
        \end{tabular}
    \end{threeparttable}
\end{table}

\subject{Model poisoning attacks under the default FL setting.} As mentioned earlier, when it comes to model poisoning attacks, we consider it is only practical for the cross-device FL but not the cross-silo FL, as it tends to be unlikely for an attacker to compromise an organization participating in the cross-silo FL. 
When evaluating model poisoning, different fractions of compromised clients were experimented and they are $M \in \{1\%, 5\%, 10\%, 20\%, 30\%, 40\%\}$. 
However, unlike the data poisoning where $M \leq 5\%$ are considered as practical, only $M \leq 1\%$ is considered as practical since model poisoning requires a higher threat bar than data poisoning in the real world. 
Regarding to the specific model poisoning algorithms, we consider three representative ones: LIE (little is enough)~\cite{baruch2019little}, MIN-MAX~\cite{shejwalkar2021manipulating}, and MIN-SUM~\cite{shejwalkar2021manipulating}, all of which are AGR-agnostic and thus qualify our threat model. 

Across these algorithms, we assume the attackers have no access to benign FL clients (and their updates), but full access to the original benign updates of the compromised FL clients. 
In a nutshell, the LIE attack adds to each dimension of the model updates of the compromised clients, a perturbation that is small enough to not be detected and pruned as outliers by robust AGRs. Unlike LIE, MIN-MAX and MIN-SUM formulate the perturbation generation as an optimization problem wherein the resulting malicious model update is upper bounded by its distance to benign updates of compromised FL clients. Particularly, In MIN-MAX, the distance of the resulting malicious model update to any benign updates is upper bounded by the maximum distance between any pair of benign updates, whereas MIN-SUM requires the sum of squared distance between the malicious update and all benign updates should be no more than that of all pairs of benign updates. For more details regarding these attack algorithms, please refer to the original papers~\cite{baruch2019little,shejwalkar2021manipulating}.

Table~\ref{tab:pratical_model_poison_default_fl} lists the attack impact of these algorithms in terms of the decrease in model accuracy. We can see that, regardless of the attack techniques, the practical model poisoning attack ($M \leq 1\%$) has a negligible impact on accuracy loss (less than 0.2\% ).  
One exception resides in the attacks using MIN-MAX and MIN-SUM against the spam detection model, which have achieved a loss in accuracy by 1.52\% and 1.30\% respectively,  which is still minor and thus acceptable, considering the extra benefits of FL. 
Then, even if the $M$ is set at an impractically high value, e.g., 30\%, the attack impact is still very minor for Android malware detection (with the highest impact achieved by MIN-SUM at 0.53\%). We also see some negative but minor values for the attack impact, which were caused by the inherent randomness of our experiments and suggest the respective model poisoning attacks have no impact on the model performance.

\begin{table}
    \centering
    \footnotesize
    \caption{The attack impact of practical \textbf{model poisoning attacks}  under label-based non-IID data distribution.}
    \label{tab:pratical_model_poison_label_non_iid}
    \begin{threeparttable}
        \begin{tabular}{cccccc}
            \toprule
            Task & $M$ & Non-IID & LIE & MIN-MAX & MIN-SUM \\
            \midrule
             \multirow{8}{*}{Spam}
                & \multirow{4}{*}{1\%} & 0.5 &$-0.0025$ & $0.0131$& $0.0107$\\
                & & 1& $-0.0008$ & $0.0199$& $0.0287$\\
                & & 5&$-0.0004$ & $0.0267$ & $0.0282$ \\
                & & 10 &$-0.0019$ &$ 0.0250$ & $0.0223$\\
                \cline{2-6}
                & \multirow{4}{*}{5\%} & 0.5 & $-0.0041$ & $0.0213$ & $0.0240$\\
                & &  1 & $-0.0007$ & $0.0338$ &$0.0329$\\
                & &  5 & $0.0015$ & $0.0485$ & $0.0451$\\
                & &  10 & $-0.0018$ & $0.0433$ & $0.0407$\\
            \midrule
            \multirow{8}{*}{Malware}
                & \multirow{4}{*}{1\%} &  0.5 & $0.0031$ & $0.0086$ & $0.0033$\\
                & &  1 & $0.0015$ & $0.0072$ & $0.0054$\\
                & &  5 & $0.0027$ & $-0.0026$ & $-0.0071$\\
                & &  10 & $-0.0022$ & $0.0053$ & $0.0032$\\
            \cline{2-6}
                & \multirow{4}{*}{5\%} & 0.5 & $0.0045$ & $-0.0060$ & $-0.0011$ \\
                & &  1 & $0.0055$ & $0.0067$ & $0.0052$\\
                & &  5 & $-0.0052$ & $-0.0025$ & $-0.0022$\\
                & &  10 & $-0.0040$ & $0.0015$ & $-0.0031$\\
            
           \bottomrule
        \end{tabular}
    \end{threeparttable}
\end{table}

\subject{Model poisoning attacks under label-based non-IID distribution.} As 
FL clients in realistic security training scenarios tend to be subject to label-based non-IID data distributions, we then moved on to evaluate model poisoning attacks under different label-based non-IID data distributions. Again, the Dirichlet distribution is used to profile the extent of label-based non-IID and various $\alpha$ values were considered: $\{0.5, 1, 5, 10\}$. 
Table~\ref{tab:pratical_model_poison_label_non_iid} presents the attack impact of model poisoning under different attack settings of $M \in \{1\%, 5\%\}$. 
Similar to the default FL scenario, practical model poisoning attacks ($M \leq 1\%$) under label-based non-IID scenarios have incurred a very constrained impact on security detection tasks. Specifically, on one hand, none of the three model poisoning attacks have any impact on the Android malware detection task. On the other hand, for SMS spam detection, the MIN-MAX attack has led to the maximum decrease of accuracy by 2.50\%, which however is still not much.

\begin{table}
    \centering
    \footnotesize
    \caption{The effect of different AGRs in terms of defending against \textbf{model poisoning} attacks. Across these experiments, the FL clients are subject to a label-based non-IID Dirichlet distribution with $\alpha = 0.5$.}
    \label{tab:model_poison_robust_agr}
        \begin{threeparttable}
        \begin{tabular}{p{0.08\textwidth}lcc}
            \toprule
           Setting & 
          AGR & MIN-MAX & MIN-SUM\\
            \midrule
            \multirow{3}{0.08\textwidth}{Spam with $M = 1\%$}
                & FedAVG  & $0.0131 $& $0.0107$\\
                & Trimmed Mean&$0.0014$&$0.0009$ \\
                &Multi-Krum &$0.0028$&$0.0077$\\
           \bottomrule
        \end{tabular}
    \end{threeparttable}
\end{table}

\subject{Defending against effective model poisoning attacks via robust AGRs.} 
As some practical model poisoning attacks have resulted in a minor performance degradation for the SMS spam detection task, we further explored the effect of robust AGRs on defending against such model poisoning attacks. Particularly, two robust AGRs (Trimmed Mean and Multi-Krum) were selected under our experiments. As shown in Table~\ref{tab:model_poison_robust_agr},  both Trimmed Mean and Multi-krum work well in terms of  decreasing the attack impact. Particularly, Trimmed Mean has decreased the attack impact of MIN-MAX from 1.31\% to 0.14\% while the attack impact of MIN-SUM has decreased from 1.07\% to 0.09\%.

\subject{Model poisoning with impractically high fractions of compromised FL clients under label-based non-IID distribution.} 
Furthermore, we also evaluated impractical fractions of compromised FL clients with $M \in \{10\%, 20\%, 30\%, 40\%\}$, and the detailed results are presented in Appendix~\ref{appendix:impractical_model_poisoning}. The key messages we can distill from these results are two-fold. On one hand, model poisoning attacks under impractically high $M$ values can lead to not only significant performance degradation, but also even convergence failures. For instance, when the fraction of compromised FL clients is $40\%$, the MIN-SUM model poisoning attack can degrade the SMS spam detection by 7.56\% in accuracy as well as causing convergence failures for the Android malware detection task.   On the other hand, the defensive effect of robust AGRs varies significantly across different $M$ settings and security prediction tasks, which calls for task-specific and scenario-specific prior evaluations before deploying robust AGRs. Particularly, for some attacking scenarios (e.g., MIN-SUM attack at $M = 40\%$), robust AGRs can decrease the attack impact from over 45\% to only 1\%, while in some other attacking settings, they can have no defensive effect or even increase the attack impact.

%% file: 5_2_fl_efficiency.tex
In this section, we summarize the efficiency issues as observed in our FL experiments, and highlight a few promising strategies to address these issues. 

\begin{table}
    \centering
    \footnotesize
    \caption{The performance of threat detection models that are FL-trained with different AGRs.}
    \label{tab:efficiency_accuracy_AGR}
    \begin{threeparttable}
        \begin{tabular}{ccccc}
            \toprule
            Task & Scenario & Trimmed Mean & Multi-Krum  & FedAVG \\
            \midrule
            \multirow{1}{*}{Spam}
                & Cross-device & 97.47\% &98.03\% &98.32\% \\
            \multirow{1}{*}{Malware}
                & Cross-device & 97.34\% & 97.74\% & 98.49\% \\
           \bottomrule
        \end{tabular}
    \end{threeparttable}
    
\end{table}

\begin{figure}
    \centering
    \subfigure[Spam in cross-device FL.]{
        \label{fig:efficiency_AGR_spam_device}
        \includegraphics[width=.45\columnwidth]{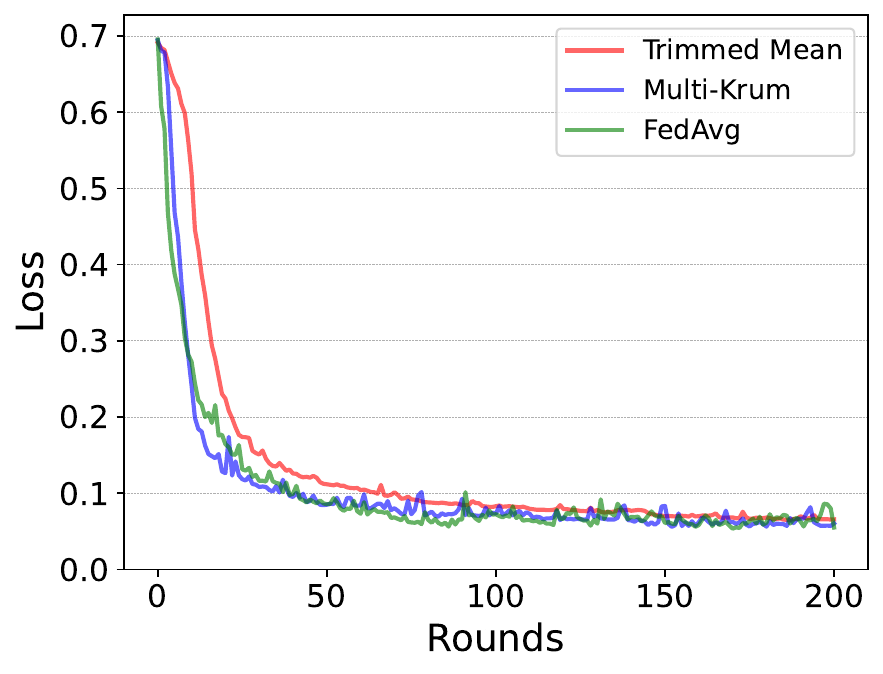}
    }
    \subfigure[Malware in cross-device FL.]{
        \label{fig:efficiency_AGR_malware_device}
        \includegraphics[width=.45\columnwidth]{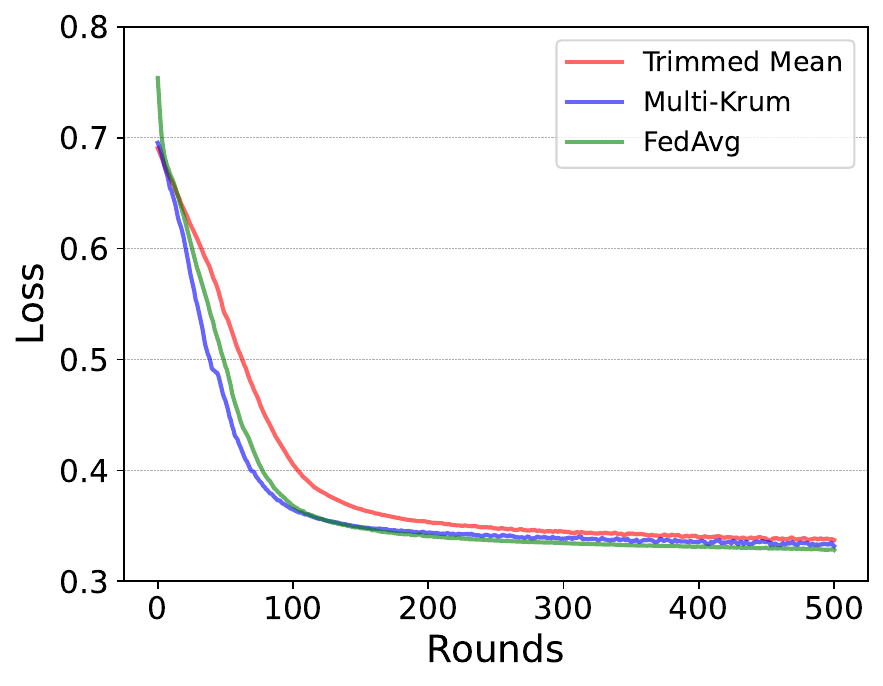}
    }
    \caption{The FL training processes of different AGRs in the default cross-device FL settings. }
    \label{fig:efficiency_AGR}
\end{figure}

\begin{table}[t]
    \centering
    \footnotesize
    \caption{The convergence of robust AGRs for the default FL settings.}
    \label{tab:efficiency_AGR}
    \begin{threeparttable}
        \begin{tabular}{cccccc}
            \toprule
            \multirow{2}{*}{Task} & \multirow{2}{*}{Scenario} & \multicolumn{2}{c}{Trimmed Mean} & \multicolumn{2}{c}{Multi-Krum} \\
         & & Rounds\tnote{1} &  Time\tnote{2} & Rounds & Time  \\
            \midrule
            \multirow{1}{*}{Spam}
                 & Cross-device & 218.18\% & 195.61\% & 101.52\% & 121.58\%\\
            \multirow{1}{*}{Malware}
                 & Cross-device & 122.58\% & 118.73\% & 79.03\% & 75.75\%\\ 
           \bottomrule
        \end{tabular}
        \begin{tablenotes}
            \item [1] The ratio of the number of FL rounds over that of FedAVG.
            \item [2] The ratio of time cost for the respective robust AGR over that of FedAVG.
        \end{tablenotes}
    \end{threeparttable}
    
\end{table}

\finding{When enabled by default for FL training, Multi-Krum, a robust AGR, has a minor overhead for the model performance and the convergence time.}

\subject{The (in)efficiency of robust AGRs.}
As revealed by our byzantine resilience evaluation, robust AGRs can help mitigate attacks of both model poisoning and data poisoning to some extent in cross-device FL. However, it is unclear whether such a benefit makes it worth to enable robust AGRs by default in a routine FL training process, since model poisoning attacks may not always be present but efficiency matters a lot in FL training. We thus moved to profile the overheads of aforementioned two robust AGRs along with a direct comparison with FedAVG under the default FL settings. Table~\ref{tab:efficiency_accuracy_AGR} presents the accuracy stats of the resulting models and we can see enabling robust AGRs have a minor performance impact. 

We then looked into the training processes and compared different AGRs with regards to the training time and the number of convergence rounds. We ran each settings for three times and obtained the average.
As shown in Figure~\ref{fig:efficiency_AGR} and Table~\ref{tab:efficiency_AGR}, Trimmed Mean has led to both a \textit{slow} training process and a notable increase in training time. When compared with FedAVG, it increased the number of cross-device FL training rounds by 118.18\% and 22.58\% for SMS spam detection and Android malware detection respectively. Similarly, the time cost also increased by 95.61\% and 18.73\%. On the other hand, the overheads of Multi-Krum appear to be minor across the two security prediction tasks. For SMS spam detection, Multi-Krum has increased the time of cross-device FL training by only 21.58\% while barely increasing the rounds needed to converge. However, for Android malware detection, it has made the training process more efficient along with fewer training rounds and training time. A likely explanation is that our SMS spam detection model is much larger than the Android malware detection model, and the large-volume parameters can make it costly to calculate Euclidean distance between model updates, which however is necessary in Multi-Krum when selecting out qualified model updates. Based on these results, we conclude that Multi-Krum can be enabled by default in cross-device FL training so as to defend possible poisoning attacks while incurring few or even no overheads, especially when the model under training is of a small size.

\finding{A label-based non-IID data distribution can incur a significant delay in convergence, which however can be effectively mitigated through the bootstrapping strategy.}

\begin{figure}
    \centering
    \includegraphics[width=.8\columnwidth]{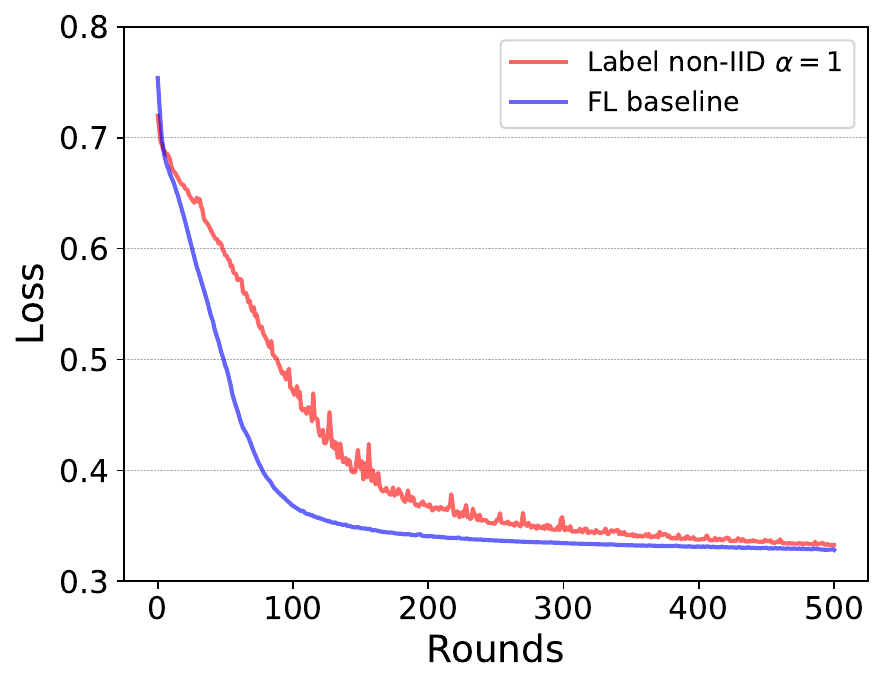}
    \caption{The convergence processes of the baseline model (quantity-based non-IID with $\alpha =1$) and the model under label-based non-IID distribution ($\alpha =1$). Both models are for Android  malware detection in cross-device FL.}
    \label{fig:fl_delay_convergence}
\end{figure}

\subject{The convergence delay due to label-based non-IID data distribution.} What is also observed is that extreme label-based non-IID data distributions can lead to notable convergence delay in cross-device FL. 
For instance, as illustrated in Figure~\ref{fig:fl_delay_convergence} for the task of Android malware detection, the label-based non-IID distribution with $\alpha = 1$ has increased the cross-device FL training process by 200 more rounds when compared to the training process under the default FL setting. 

\subject{Mitigating the convergence delay through bootstrapping.}
\label{sub:bootstrapping}
We consider a bootstrapping strategy to address the aforementioned additional delays caused by non-IID data distributions. In a nutshell, our bootstrapping strategy first pre-trains an initial model using server-side publicly available datasets (e.g., open source Android malware datasets), and then fine-tunes the pre-trained model using local datasets from FL clients. Similar ideas have been widely considered in various fields of FL, e.g, personalised FL~\cite{Fallah2020}.

\begin{figure}
    \centering
    \subfigure[Cross-device FL with 200 clients.]{
        \label{fig:fl_bootstrapping_200}
        \includegraphics[width=.45\columnwidth]{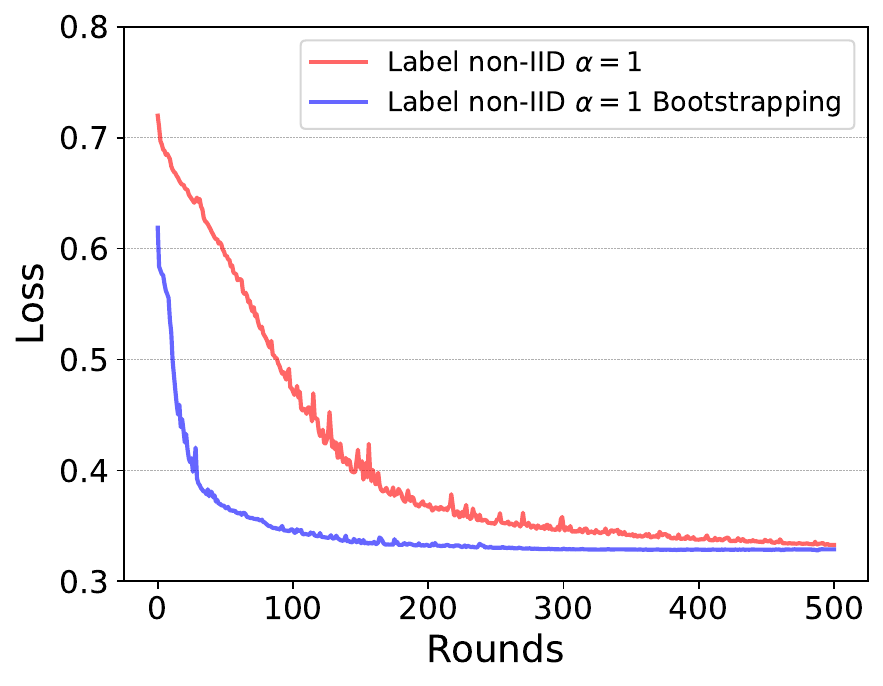}
    }
    \subfigure[Cross-device FL with 500 clients.]{
        \label{fig:fl_bootstrapping_500}
        \includegraphics[width=.45\columnwidth]{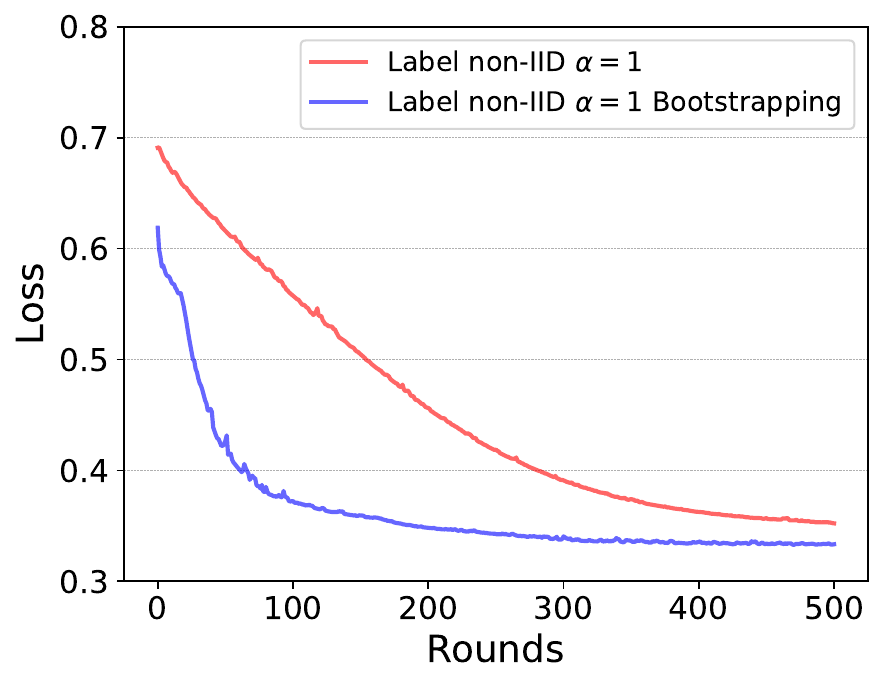}
    }
    \caption{The convergence processes for Android malware detection models trained with or without bootstrapping. Label non-IID $\alpha = 1$ denotes the label-based non-IID distribution with $\alpha = 1$.}
    \label{fig:fl_bootstrapping}
\end{figure}

To evaluate the effectiveness of this bootstrapping strategy, we first trained a CNN-based Android malware detection model using the Genome Android malware dataset in the central server to simulate the pre-training step. This pre-trained model achieved $98.67\%$ in accuracy, $99.23\%$ in precision and $98.47\%$ in recall when evaluated on the test part of the Genome dataset  at the small cost of training for 30 rounds. We then tested the model on our held-out testing dataset that is default for the Android malware detection. It gained only $68.71\%$ in accuracy, $95.95\%$ in precision and $39.06\%$ in recall. We then considered this pre-trained but weak model as the initial global model in the FL scenario, and applied FL training to it, using a label-based non-IID distribution with $\alpha =1$. 
In addition to trying the default FL setting of 200 clients, we also tested a setting with 500 clients. The convergence process is shown in Figure~\ref{fig:fl_bootstrapping}, along with a comparison to their respective non-bootstrapped counterparts. We can see that the bootstrapping settings have effectively improved not only the convergence speed but also the stability in label-based non-IID scenarios. 
These results suggest the effectiveness of using bootstrapping in FL to speed up and stabilize the convergence of FL training, especially for extreme non-IID data distributions.

%% file: 6_discuss.tex
\subject{Limitations and Future works.} In this study, we focus on evaluating the feasibility of FL for two representative and privacy-sensitive security prediction tasks, with regards to effectiveness, byzantine robustness, and efficiency. As discussed above, security tasks can still vary to a notable extent in their resilience to byzantine clients and extreme non-IID data distribution. Therefore, it is worth further evaluating more representative security prediction tasks in FL scenarios, so as to identify the case-by-case deployment strategies customized for each given security prediction task. 
On the other hand, cyber threat detection tasks can share many common properties, e.g., the threat model, the behavior patterns of the attacker, etc. We thus believe our findings and observations can likely be generalized to tasks sharing similar properties, e.g., the detection of Email/Tweet spam, and classification of malware of other platforms.

Furthermore, for both tasks of SMS spam classification and Android malware detection, we composed historical datasets from multiple sources, due to the unavailability of a threat dataset that is both up-to-date and balanced. However, as these historical datasets are either out-of-dated or imbalanced, the composed dataset may still have a different data distribution when compared with the real-world and up-to-date data. However, this issue appears to be unavoidable and its impact on our FL evaluations may deserve further investigation.

Besides, when evaluating the byzantine resilience, we focus on untargeted poisoning attacks since it is more challenging and can be more destructive when compared with targeted poisoning attacks. However, targeted poisoning attacks may still occur for FL-based security prediction tasks and we leave it as our future works to further explore the byzantine resilience of FL against targeted poisoning attacks. 
Then, when it comes to the efficiency of FL, we focus on the computational aspect, and leave it as a future work to evaluate the communication efficiency of FL-based security prediction tasks, especially in realistic deployment settings.

\subject{Data and code release.} 
We released the source code for reproducing all aforementioned FL experiments along with the respective datasets. The repository link is \url{https://github.com/ChaseSecurity/Fostering_Cyber_Threat_Detection_Through_FL}.

%% file: conclusion.tex
In this study, we have extensively explored the feasibility of applying federated learning to  cyber threat detection tasks. Our evaluations have demonstrated that FL is both effective and byzantine-resilient when applied to representative threat detection tasks, and its efficiency issues can be well addressed through a bootstrapping strategy. To conclude, FL is a promising option for enabling privacy-preserving cyber threat detection.  

%% file: 9_appendix.tex
\section{More Results on the Effectiveness of FL}

\subsection{More performance results in the consistent label imbalance Scenarios.}
\label{appendix:cli_results}

Table~\ref{tab:cli_precision} presents the precision performance for FL-trained models in consistent label imbalance scenarios while Table~\ref{tab:cli_recall} presents the recall performance.

\begin{table}[h]
    \centering
    \footnotesize
    \caption{The precision stats of security models under consistent label imbalance.}
    \label{tab:cli_precision}
    \begin{threeparttable}
        \begin{tabular}{ccccc}
            \toprule
            \multirow{2}{*}{Task} & \multirow{2}{*}{Model} & \multicolumn{3}{c}{$PNR$}\\
            && 0.25& 1 & 4\\
            \midrule
            \multirow{3}{*}{Spam} 
            & Central & $0.9889$&$0.9912$ &$0.9842$ \\
            &Cross-Device FL & $0.9930$&$0.9884$ &$0.9454$ \\
            &Cross-Silo FL & $0.9952$& $0.9923$&  $0.9722$\\
            \midrule
            \multirow{3}{*}{Malware} 
            & Central & $0.9703 $& $0.9779$ & $0.7866$ \\
            &Cross-Device FL & $0.9742$ & $0.9837$ & $0.5000$ \\
            &Cross-Silo FL & $0.9810$ & $0.9846$ & $0.5013$ \\
            \bottomrule
        \end{tabular}
    \end{threeparttable}
\end{table}

\begin{table}[h]
    \centering
    \footnotesize
    \caption{The recall stats of security models under consistent label imbalance.}
    \label{tab:cli_recall}
    \begin{threeparttable}
        \begin{tabular}{ccccc}
            \toprule
            \multirow{2}{*}{Task} & \multirow{2}{*}{Model} & \multicolumn{3}{c}{$PNR$}\\
            && 0.25& 1 & 4\\
            \midrule
            \multirow{3}{*}{Spam} 
            & Central &$0.9841$ &$0.9882$ &$0.9905$ \\
            &Cross-Device FL & $0.9474$&$0.9829$ & $0.9852$\\
            &Cross-Silo FL & $0.9762$& $0.9886$&$0.9926$ \\
            \midrule
            \multirow{3}{*}{Malware} 
            & Central & $0.9608$ & $0.9948$ & $1.0000$ \\
            &Cross-Device FL & $0.9871$ & $0.9967$ & $1.0000$ \\
            &Cross-Silo FL & $0.9699$ & $0.9944$ & $1.0000$ \\
            \bottomrule
        \end{tabular}
    \end{threeparttable}
\end{table}

\subsection{Metrics to profile the stability of FL convergence in non-IID scenarios}
\label{appendix:convergence_instability}

Given the sequence of models trained during the last 10 rounds of a FL training process, let's define $\text{ACC}_{STD}$ as the standard deviation for their respective accuracy values, $\text{PRE}_{STD}$ for the precision metric, and $\text{REC}_{STD}$ for the recall metric. Table~\ref{tab:fl_convergence} lists the values of these metrics for all the aforementioned non-IID scenarios. In Figure~\ref{fig:fl_unstable_convergence}, we can see that the whole converging process is unstable specially in the cross-device FL. It further strengthens our observation that non-IID distributions can lead to higher instability of the FL training process in terms of convergence.

\begin{table}[h]
    \centering
    \footnotesize
    \caption{The (in)stability of the convergence of FL training.}
    \label{tab:fl_convergence}
    \begin{tabular}{p{0.04\textwidth}ccccc}
        \toprule
         Task& Non-IID & Params &  $\text{ACC}_{STD}$ & $\text{PRE}_{STD}$ & $\text{REC}_{STD}$\\
         \midrule
         \multirow{4}{0.04\textwidth}{Spam} & Default & $\alpha = 1$&$0.0007$&$0.0042$&$0.003$6 \\
         & Quantity & $\alpha = 0.5$ &$0.0009$&$0.0037$&$0.0024$\\
         &  Label & $\alpha = 0.5$ &$0.0081$&$0.0201$&$0.0116$\\
         & CLI & $PNR = \frac{1}{4}$ &$0.0034$&$0.0036$&$0.101$\\
         \midrule 
        \multirow{4}{0.04\textwidth}{Malware} & Default & $\alpha = 1$ & $0.0005$ & $0.0007$ & $0.0011$ \\
         & Quantity & $\alpha = 0.5$ & $0.0012$ & $0.0022$ & $0.0000$\\
         &  Label & $\alpha = 0.5$ & $0.0012$ & $0.0021$ & $0.0000$\\
         & CLI & $PNR = \frac{1}{4}$ & $0.0008$ & $0.0016$ & $0.0016$\\
         \bottomrule
    \end{tabular}
\end{table}

\section{More Results of Data Poisoning Attacks}

\subsection{Data poisoning attacks in label-based non-IID scenarios}
\label{appendix:data_poison_label_non_idd}
Table~\ref{tab:pratical_data_poison_label_non_iid} presents the attack impact of data poisoning attacks in various label-based non-IID scenarios.

\begin{table*}
    \centering
    \footnotesize
    \caption{The attacking impact in accuracy decrease for \textbf{data poisoning  attacks} under label-based non-IID distribution.}
    \label{tab:pratical_data_poison_label_non_iid}
    \begin{threeparttable}
        \begin{tabular}{cccccc}
            \toprule
                       
            \multirow{2}{*}{Poisoning Setting} & 
            \multirow{2}{*}{Non-IID}
                & \multicolumn{2}{c}{Spam} & \multicolumn{2}{c}{Malware}\\
                && Device & Silo & Device & Silo\\
            \midrule
            \multirow{5}{0.2\linewidth}{$M = 5\%, p = 50\%$}
                & Balanced &$0.0013\pm0.0008$&$-0.0002\pm0.0014$ & $0.0093\pm0.0008$ & $0.0044\pm0.0017$\\
                & $\alpha = 0.5$& $-0.0024\pm0.0072$& $-0.0001\pm0.0009$ & $0.0020\pm0.0024$ & $0.0031\pm0.0020$ \\
                & $\alpha = 1$ &$-0.0009\pm0.0047$& $-0.0003\pm0.0007$ & $0.0006\pm0.0033$ & $-0.0012\pm0.0024$\\
                & $\alpha = 5$ & $-0.0002\pm0.0012$&$-0.0001\pm0.0005$ & $0.0073\pm0.0009$ & $0.0002\pm0.0028$\\
                & $\alpha = 10$ & $-0.0014\pm0.0012$&$0.0002\pm0.0006$ & $0.0056\pm0.0009$ & $-0.0109\pm0.0013$\\
             \midrule
             \multirow{5}{0.2\linewidth}{$M = 5\%, p = 100\%$} 
                & Balanced & $0.0023\pm0.0022$&$-0.0007\pm0.0010$ & $0.0020\pm0.0012 $ & $0.0007\pm0.0011$\\
                & $\alpha = 0.5$ &$-0.0018\pm0.0092$ &$0.0006\pm0.0007$ & $0.0013\pm0.0037$ & $0.0134\pm0.0014$\\
                & $\alpha = 1$ & $-0.0001\pm0.0042$&$0.0013\pm0.0009$ & $-0.0014\pm0.0040$ & $-0.0052\pm0.0028$\\
                & $\alpha = 5$ & $0.0019\pm0.0018$& $-0.0004\pm0.0010$& $0.0096\pm0.0010$ & $-0.0011\pm0.0019$\\
                & $\alpha = 10$ & $-0.0012\pm0.0014$& $-0.0007\pm0.0008$& $0.0008\pm0.0016$ & $0.0000\pm0.0016$\\
           \bottomrule
        \end{tabular}
    \end{threeparttable}
\end{table*}

\subsection{The FL training processes under data poisoning attacks}
\label{appendix:data_poisoning_training}
\begin{figure}
    \centering
    \subfigure[Spam in cross-device FL.]{
        \label{fig:fl_data_poisoning_default_M_setting_spam_cross_device}
        \includegraphics[width=.45\columnwidth]{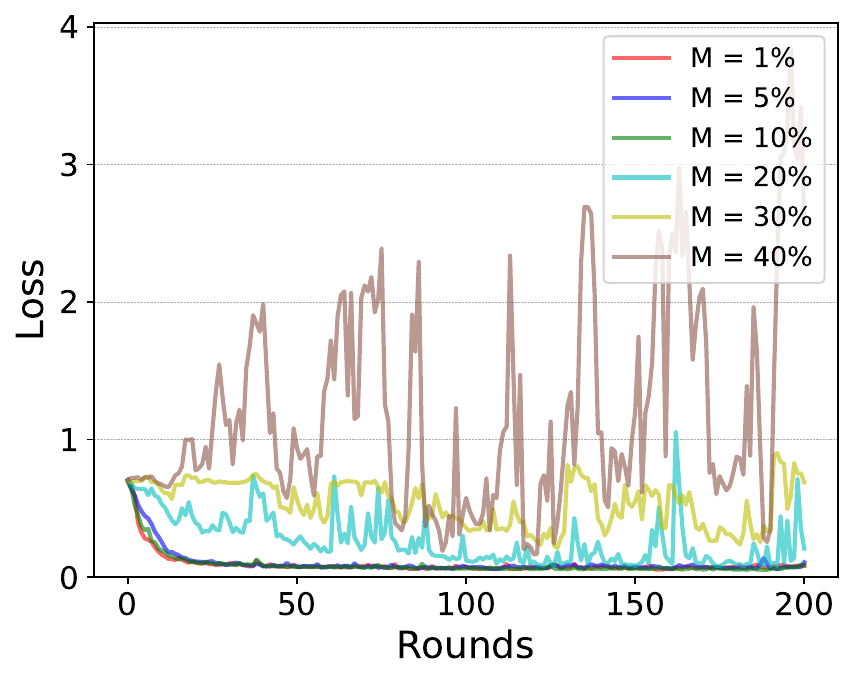}
    }
    \hfill
    \subfigure[Spam in cross-silo FL.]{
        \label{fig:fl_data_poisoning_default_M_setting_spam_cross_silo}
        \includegraphics[width=.45\columnwidth]{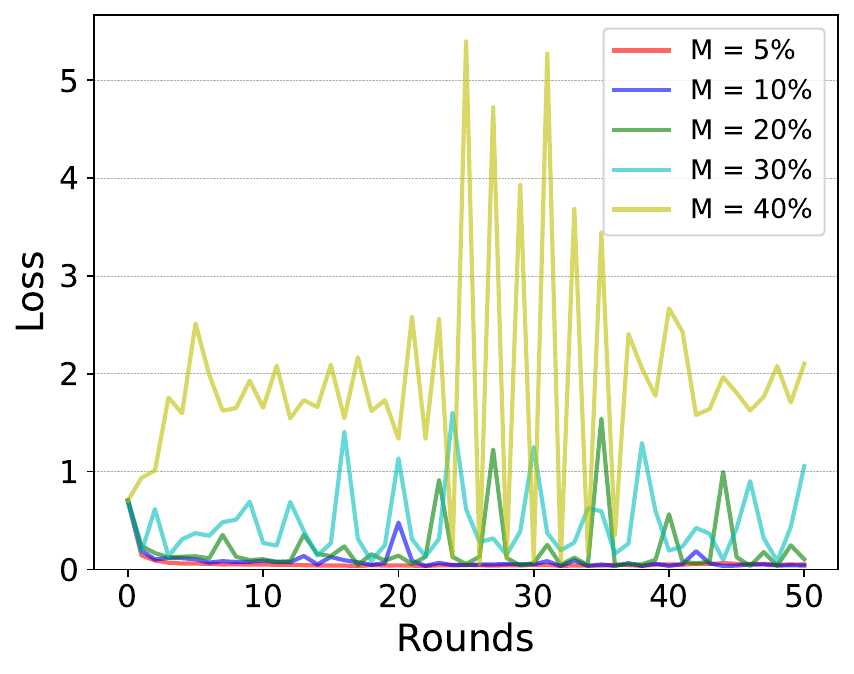}
    }
    \subfigure[Malware in cross-device FL.]{
        \label{fig:fl_data_poisoning_default_M_setting_malware_cross_device}
        \includegraphics[width=.45\columnwidth]{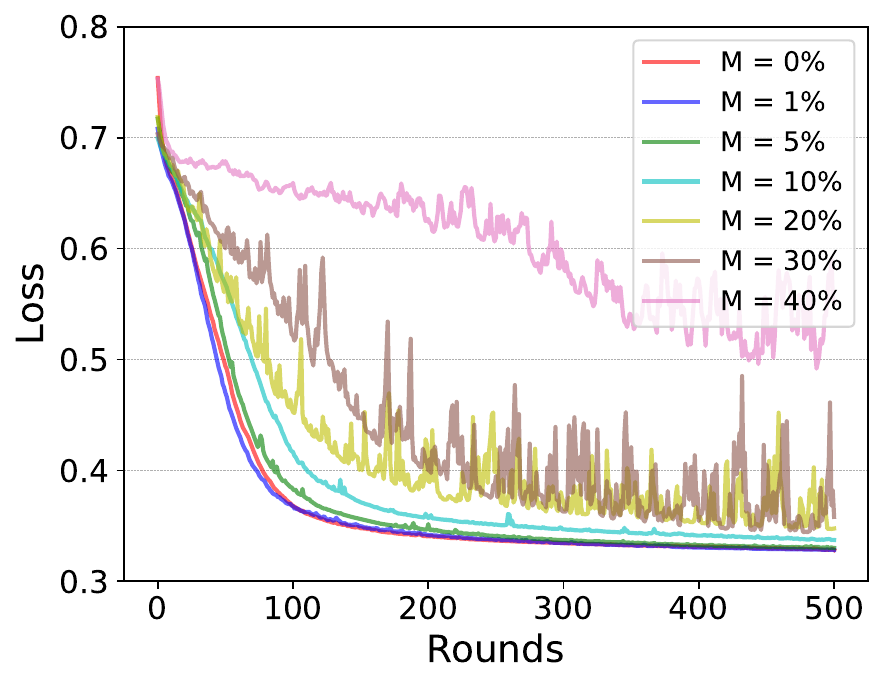}
    }
    \subfigure[Malware in cross-silo FL.]{
        \label{fig:fl_data_poisoning_default_M_setting_malware_cross_silo}
        \includegraphics[width=.45\columnwidth]{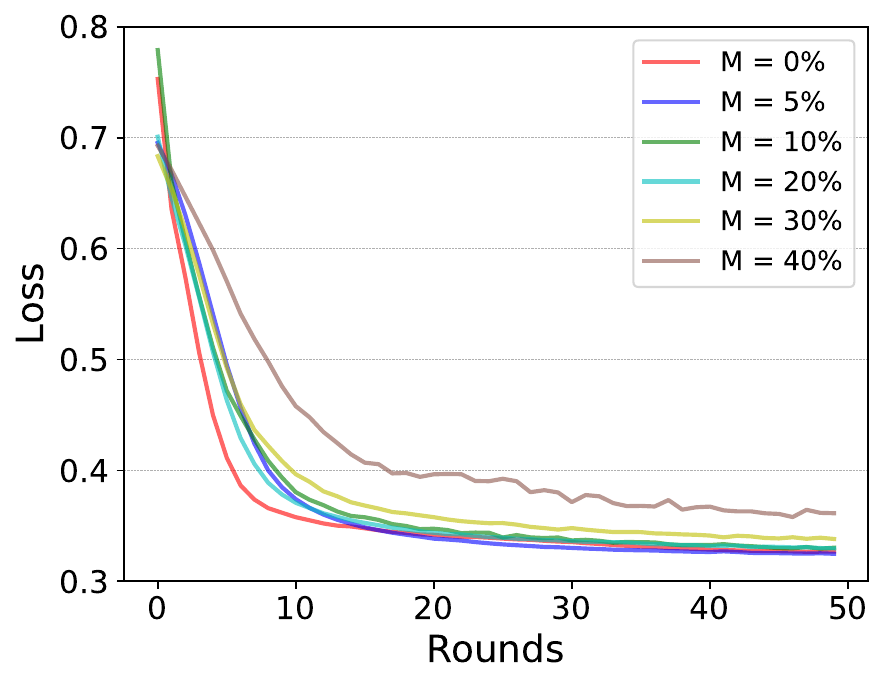}
    }
    \caption{The training process of data poisoning via label flipping for the default FL settings of $M$.}
    \label{fig:fl_data_poisoning_default_M_setting}
\end{figure}

Figure~\ref{fig:fl_data_poisoning_default_M_setting} presents the training process for security detection models under data poisoning attacks. 

\begin{figure}[h]
    \centering
    \subfigure[Spam in cross-device FL.]{
        \label{fig:fl_data_poisoning_default_p_setting_spam_cross_device}
        \includegraphics[width=.45\columnwidth]{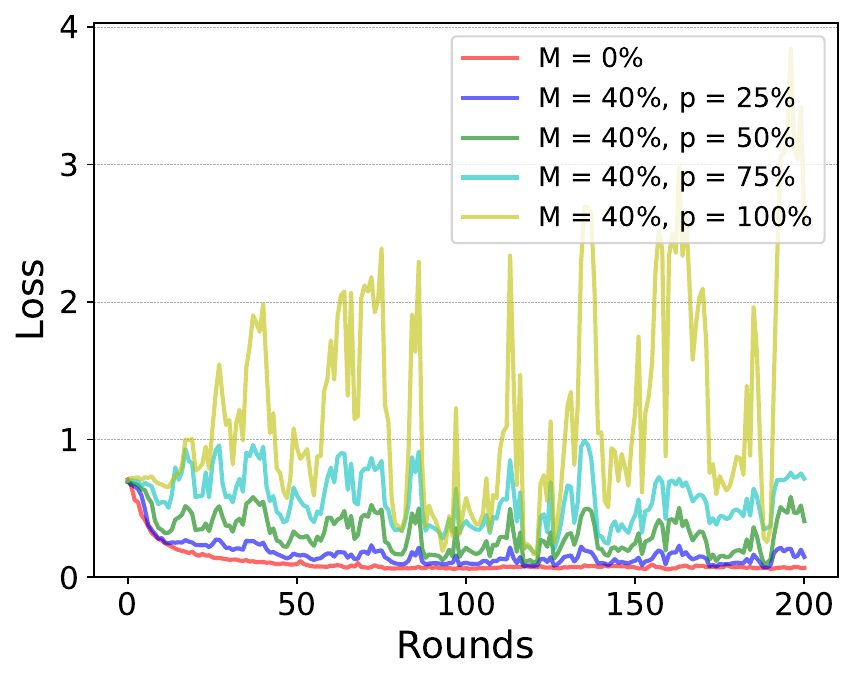}
     }
    \subfigure[Malware in cross-device FL.]{
        \label{fig:fl_data_poisoning_default_p_setting_malware_cross_device}
        \includegraphics[width=.45\columnwidth]{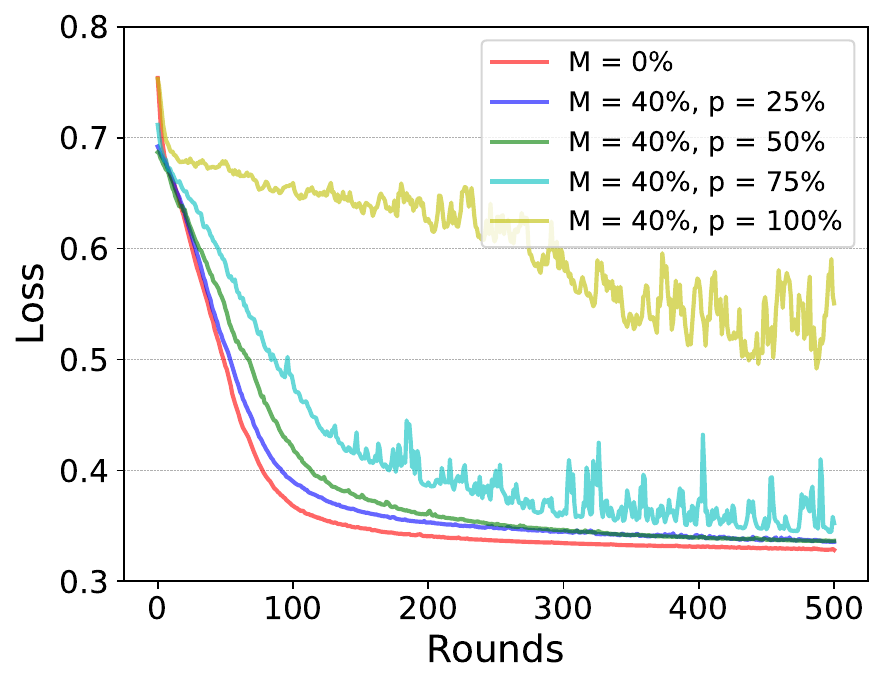}
    }

    \caption{The training process of data poisoning via label flipping for the default FL settings of $p$. }
    \label{fig:fl_data_poisoning_default_p_setting}
\end{figure}

When $M = 40\%$, we further explored the effect of different sample poisoning rates $p$, and Figure~\ref{fig:fl_data_poisoning_default_p_setting} presents the respective training processes for security detection models under such data poisoning attacks.

\section{Model Poisoning Under Impractical Settings}
\label{appendix:impractical_model_poisoning}

\begin{table}[h]
    \centering
    \footnotesize
    \caption{The attacking impact of \textbf{model poisoning attacks} with high but impractical $M$.}
    \label{tab:model_poison_label_non_iid_high_m}
    \begin{threeparttable}
        \begin{tabular}{cccccc}
            \toprule
            Task & $M$ & Non-IID & LIE & MIN-MAX & MIN-SUM \\
            \midrule
            \multirow{16}{*}{Spam}
                & \multirow{4}{*}{10\%} & 0.5 & -0.0020&0.0231 & 0.0226\\
                & &  1 &0.0003 &0.0314 &0.0397\\
                & &  5 & 0.0011&0.0745 &0.0632\\
                & &  10 & 0.0003& 0.0672&0.0669\\
                \cline{2-6}
                & \multirow{4}{*}{20\%} & 0.5 & -0.0033& 0.0281&0.0196\\
                & &  1 &-0.0000&0.0456 &0.0387 \\
                & &  5 & -0.0002& 0.0400&0.0585\\
                & &  10 &-0.0009 &0.0757 &0.0662\\
               \cline{2-6}
                & \multirow{4}{*}{30\%} & 0.5 &-0.0009 &0.0378 &0.0365\\
                & &  1 &-0.0020 & 0.0320&0.0344\\
                & &  5 &0.0018 &0.0812 &0.0617\\
                & &  10 &-0.0001 &0.0457 &0.0649\\
                \cline{2-6}
                & \multirow{4}{*}{40\%} & 0.5 &-0.0005 & 0.0396&0.0311\\
                & &  1 & 0.0022& 0.0452&0.0450\\
                & &  5 & 0.0035&0.0478 &0.0555\\
                & &  10 &-0.0007 &0.0725 &0.0756\\                
            \midrule
            \multirow{16}{*}{Malware}
                & \multirow{4}{*}{10\%} & 0.5 & $0.0056$ & $0.0027$ & $-0.0054$ \\
                & &  1 & $0.0039$ & $0.0089$ & $0.0062$ \\
                & &  5 & $-0.0015$ & $-0.0055$ & $-0.0074$\\
                & &  10 & $0.0001$ & $-0.0002$ & $-0.0024$ \\
                \cline{2-6}
                & \multirow{4}{*}{20\%} & 0.5 & $0.0072$ & $0.0205$ & $0.0145$ \\
                & &  1 & $0.0067$ & $0.0134$ & $0.0115$ \\
                & &  5 & $0.0001$ & $0.0024$ & $0.0068$\\
                & &  10 & $-0.0028$ & $0.0045$ & $-0.0038$ \\
                \cline{2-6}
                & \multirow{4}{*}{30\%} & 0.5 & $0.0046$ & $0.1025$ & $0.3148$ \\
                & &  1 & $0.0079$ & $0.0262$ & $0.1964$ \\
                & &  5 & $-0.0016$ & $0.0101$ & $0.0081$ \\
                & &  10 & $-0.0012$ & $0.0080$ & $0.0059$ \\
                \cline{2-6}
                & \multirow{4}{*}{40\%} & 0.5 & $0.0106$ & $0.4312$ & $0.4760$ \\
                & &  1 & $0.0041$ & $0.4639$ & $0.4551$ \\
                & &  5 & $0.0009$ & $0.0242$ & $0.4689$ \\
                & &  10 & $0.0019$ & $0.0088$ & $0.0236$\\
            
           \bottomrule
        \end{tabular}
    \end{threeparttable}
\end{table}

Table~\ref{tab:model_poison_label_non_iid_high_m} presents the results of model poisoning attacks under high but impractical attacking settings, i.e., $M \in \{10\%, 20\%, 30\%, 40\%\}$. As we can see, the higher the fraction of compromised FL clients ($M$), the more the attack impact is, which applies to both security prediction tasks and all three attack algorithms, and thus is aligned with previous works~\cite{baruch2019little, shejwalkar2021manipulating}. Besides, among the three attack algorithms, LIE has negligible attack impact in almost all experiments except for attacking the Android malware detection task with $M = 40\%$, while MIN-MAX and MIN-SUM performed much better. 

Then, when it comes to the byzantine resilience of security prediction tasks, the two security prediction tasks vary significantly in their byzantine resilience (adversarial robustness). Particularly, when $M \leq 20\%$, the task of Android malware detection tends to be more resistant against model poisoning attacks than SMS spam detection. However, as $M$ continues to increase, the malware detection model drops significantly in its performance or even fails to converge, while the SMS spam detection model becomes more resilient and suffers from less than 8\% decrease in accuracy across all attacking experiments. Such a robustness difference may be attributed to their difference in model architecture as well as the number of model parameters. Specifically, the SMS spam detection model, a transformer model, has 167 million parameters, which is much larger than the CNN model for malware detection.

\begin{figure}[h]
    \centering
    \includegraphics[width=.9\columnwidth]{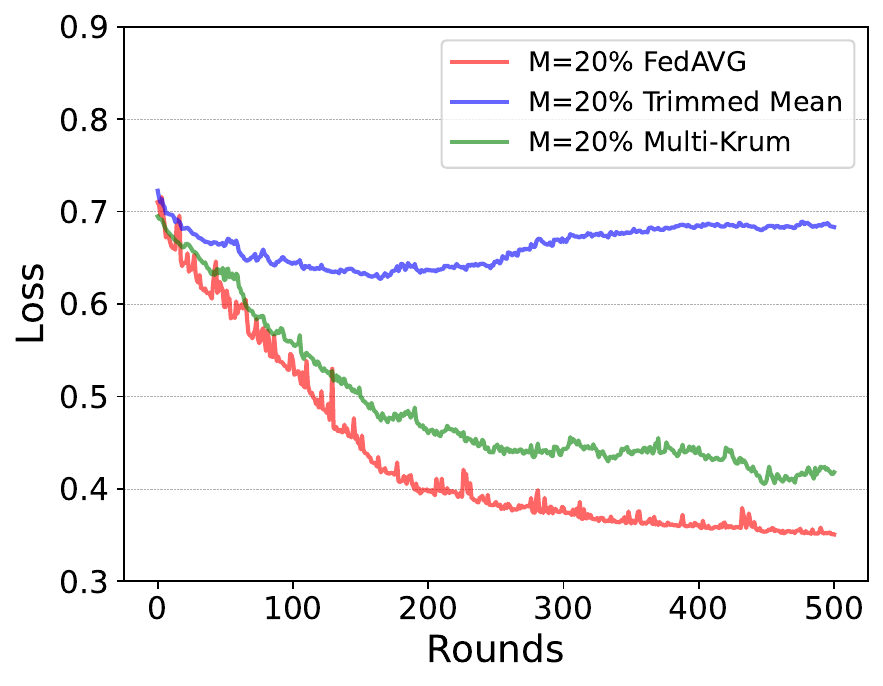}
    \caption{The FL training processes with or without the enforcement of robust AGRs, for Android malware detection, under model poisoning attacks.}
    \label{fig:convergence_failures_under_robust_agrs}
\end{figure}

\begin{table}
    \centering
    \footnotesize
    \caption{The attacking impact of \textbf{model poisoning} attacks, for Android malware detection, under \textbf{different AGR algorithms} while the client-side data follows the label-based Dirichlet distribution with $\alpha = 0.5$.}
    \label{tab:impractical_model_poison_robust_agr}
        \begin{threeparttable}
        \newcommand{\firstwidth}{.15\columnwidth}
        \begin{tabular}{ccccc}
            \toprule
           $M$ & 
          AGR & LIE & MIN-MAX & MIN-SUM\\
            \midrule
            \multirow{3}{*}{10\%}
                & FedAVG & $0.0056$ & $0.0027$ & $-0.0054$\\
                & Trimmed Mean & $0.0038$ & $0.0093$ & $0.0165$\\
                &Multi-Krum & $-0.0027$ & $0.0089$ & $0.0021$ \\
            \midrule
            \multirow{3}{*}{20\%}
                & FedAVG & $0.0072$ & $0.0205$ & $0.0145$\\
                & Trimmed Mean & $0.0069$ & $0.0313$ & $0.1562$\\
                &Multi-Krum & $0.0195$ & $0.0164$ & $0.0154$\\
            \midrule
            \multirow{3}{*}{30\%}
                & FedAVG & $0.0046$ & $0.1025$ & $0.3148$\\
                & Trimmed Mean & $0.0204$ & $0.4721$ & $0.4798$\\
                & Multi-Krum & $0.0188$ & $0.4575$ & $0.0156$\\
             \midrule
            \multirow{3}{*}{40\%}
                & FedAVG & $0.0106$ & $0.4312$ & $0.4760$\\
                & Trimmed Mean & $0.0352$ & $0.4798$ & $0.4800$\\
                & Multi-Krum & $0.0347$ & $0.4760$ & $0.0105$\\
           \bottomrule
        \end{tabular}
        \begin{tablenotes}
            \item [1] Each cell denotes the decrease in model accuracy that is achieved by a model poisoning attack under a specific AGR and label-based non-IID data distribution. 
        \end{tablenotes}
    \end{threeparttable}
\end{table}

 \subject{The effect of robust AGR on mitigating model poisoning attacks with impractically high fractions of compromised FL clients.} Although above model poisoning attacks are unlikely to occur in real-world FL scenarios due to their assumption on high fractions of compromised FL clients, we still evaluated the effectiveness of multiple robust AGRs that have been claimed in previous works~\cite{xie2018generalized, NIPS2017} for defending against such model poisoning attacks. However, as listed in Table~\ref{tab:impractical_model_poison_robust_agr}, for most attacking settings on Android malware detection, neither Trimmed Mean nor Multi-Krum has notable defensive impact. Instead, as $M$ gets larger, they even intensify the impact of the respective attack, such as making the training process fail to converge, as further illustrated in Figure~\ref{fig:convergence_failures_under_robust_agrs}.